\documentclass[11pt]{article}
\usepackage{graphicx}
\usepackage{epsfig}
\usepackage{amsmath}
\usepackage{amsfonts}
\usepackage{amssymb}
\usepackage{graphicx}
\usepackage{epstopdf}
\usepackage{amsmath}
\usepackage{amssymb}
\usepackage{amsfonts}
\usepackage{amsthm}
\usepackage[usenames]{color}
\usepackage{array}

\usepackage[
      colorlinks=true,
      linkcolor=blue,
      urlcolor=blue,
      filecolor=blue,
      citecolor=red,
      pdfstartview=FitV,
      pdftitle={},
      pdfauthor={},
      pdfsubject={},
      pdfkeywords={},
      pdfpagemode=None,
      bookmarksopen=true
]{hyperref}

\usepackage{epsfig}

\usepackage{hyperref}

\def\be{\begin{equation}}
\def\ee{\end{equation}}
\def\bea{\begin{eqnarray}}
\def\eea{\end{eqnarray}}

\begin{document}

\makeatletter
\renewcommand{\theequation}{\thesection.\arabic{equation}}
\@addtoreset{equation}{section}
\makeatother

\begin{flushright}
\end{flushright}
\vspace{1cm}

\begin{centering}

  \textbf{\Large{An electrically charged\\[5pt] doubly spinning dipole black ring}}

  \vspace{0.8cm}

  {\large Jorge V. Rocha$^\natural$, Maria J. Rodriguez$^\S$, and Oscar Varela$^\Diamond$ }

  \vspace{0.5cm}

\begin{minipage}{.9\textwidth}\small \it \begin{center}
$\natural$
Centro Multidisciplinar de Astrof\'isica -- CENTRA,\\
Dept. de F\'isica, Instituto Superior T\'ecnico, Technical University of Lisbon,\\
Av. Rovisco Pais 1, 1049-001 Lisboa, Portugal \\
 {\tt jorge.v.rocha@ist.utl.pt}\\
$ \, $ \\

$\S$
Center for the Fundamental Laws of Nature, Harvard University,\\
02138 Cambridge, MA, USA\\
{\tt mjrodri@harvard.physics.edu}
\\ $ \, $ \\

$\Diamond$
Institute for Theoretical Physics and Spinoza Institute, Utrecht University, \\
3508 TD Utrecht, The Netherlands\\
{\tt o.varela@uu.nl}
    \end{center}
\end{minipage}

\end{centering}

\vspace{3cm}

\begin{abstract}

We present a new asymptotically flat, doubly spinning black ring of $D=5$ Einstein-Maxwell-dilaton theory with Kaluza-Klein dilaton coupling. Besides the mass and two angular momenta, the solution displays both electric charge and (magnetic) dipole charge. The class of solutions that are free from conical singularities is described by four parameters. We first derive the solution in six dimensions employing the inverse scattering method, thereby generalising the inverse-scattering construction by two of the current authors of Emparan's singly spinning dipole black ring. The novel black ring itself arises upon circle Kaluza-Klein reduction. We also compute the main physical properties and asymptotic charges of our new class of solutions.  Finally, we present a five-parameter generalisation of our solution.

\end{abstract}
\vfill

\thispagestyle{empty} \newpage

\tableofcontents

\setcounter{equation}{0}

\section{Introduction} \label{sec:intro}

Analysing the existence and understanding the physical properties of four- and higher-dimensional black holes can  give us important hints about gravity at both the classical and quantum levels. On the one hand,  one of the most remarkable features of classical higher-dimensional gravity is the existence of stationary black hole solutions, like the Emparan-Reall black ring \cite{Emparan:2001wn}, with no counterpart in ordinary four-dimensional General Relativity (GR). At the quantum level, on the other hand, string realisations of charged, five-dimensional black holes proved instrumental in first deriving a microscopic interpretation of the Bekenstein-Hawking entropy formula \cite{Strominger:1996sh}. While new insights on microstate counting have been more recently discovered for the modest four-dimensional Kerr black hole, both at \cite{Guica:2008mu} and close to \cite{Castro:2010fd} extremality, it is clearly of interest to extend the arena of explicit charged black hole solutions.

Considerable amount of guesswork has usually been involved in the discovery of many higher-dimensional black holes. This is for instance the case of the $D > 4$ Myers-Perry black hole \cite{Myers:1986un} or of the black ring  \cite{Emparan:2001wn} itself. Much quicker progress in the construction of new black hole examples has been possible in situations amenable to the application of solution generating techniques. In a broad sense, this is for example the case of the supersymmetric black holes that may be engineered in string theory by means of boosts and dualities, and that admit an interpretation in terms of intersecting or wrapped D-branes \cite{Strominger:1996sh,Breckenridge:1996is}. Closely related, are the supergravity techniques that exploit U-duality transformations that manifest themselves as hidden symmetries of the sigma-model lagrangians obtained upon dimensional reduction  \cite{Sen:1992ua,Cvetic:1996xz}. Another class of construction techniques builds on classification results of supersymmetric solutions in supergravity, that may be further refined to account for solutions with a horizon \cite{Elvang:2004rt,Bena:2004de,Gauntlett:2004qy}. Of course, the power of numerics has been employed to build black holes in analytically elusive situations \cite{Kleihaus:2010pr}.

Solution generating techniques have in fact allowed for the {\it design}, nearly at will, of higher-dimensional black holes. The inverse scattering method (ISM) \cite{BZ} (see \cite{Iguchi:2011qi} for a recent review), an application of which we perform in this paper, has proved extremely useful to construct solutions of vacuum gravity in cases when sufficient, but still less symmetry than in some of the above cases, is available. The first black ring solution constructed with the ISM was that of Pomeransky and Sen'kov~\cite{Pomeransky:2006bd}, generalising the Emparan-Reall black ring~\cite{Emparan:2001wn} to include a second angular momentum. Following \cite{Pomeransky:2006bd}, other (concentric) doubly spinning, asymptotically flat black rings in five dimensions were constructed by inverse-scattering methods. These include black saturns \cite{Elvang:2007rd}, di-rings  \cite{Iguchi:2007is,Evslin:2007fv} and bicycling black rings \cite{Elvang:2007hs}. More recent inverse-scattering constructions include the explicit derivation of the unbalanced Pomeransky-Sen'kov black ring \cite{Chen:2011jb,RRunp}, or rings on topologically non-trivial backgrounds \cite{Chen:2010ih,Chen:2012zb}.

The ISM  can be applied to the construction of cohomogeneity-two solutions of $D$-dimensional {\it vacuum} gravity displaying $D-2$ commuting isometries. The method essentially relies on the integrability properties of a $GL(D-2, \mathbb{R})$ non-linear sigma model on the remaining two-dimensional surface (parametrised, in Weyl coordinates, by $\rho$ and $z$ as in equation (\ref{confmetric}) below). It is thus clear why inverse-scattering techniques are so well suited to generate black rings in $D=5$ dimensions \cite{Emparan:2001wk}: these solutions do indeed admit three commuting Killing vectors, along  time and two angular directions. For the same reason, it is also apparent that the method has no possible say on the construction of vacuum black rings in dimensions greater than five (other methods, like matched asymptotic expansions, have been used to approximate the solution in these cases \cite{Emparan:2007wm}). The $D=6$ black ring, for example, has three commuting isometries, one short of the necessary number of symmetries for the ISM to work in six dimensions.

Nevertheless, as noticed by two of the present authors~\cite{Rocha:2011vv}, the ISM can still be very useful in {\it six} dimensions in order to generate asymptotically flat black ring solutions in {\it five} dimensions. More precisely,  \cite{Rocha:2011vv} provided an explicit six-dimensional inverse-scattering construction of a  dipole-charged, singly spinning black ring of $D=5$ Einstein-Maxwell-dilaton theory, with the particular dilaton coupling (equation (\ref{KKcoupling})  below) that arises from Kaluza-Klein (KK) circle reduction of $D=6$ vacuum gravity. The solution was not new: it was first constructed by inspired guesswork in \cite{Emparan:2004wy} for actually all values of the dilaton coupling, not just KK. The advantage of having pinned down a systematic, inverse-scattering construction, is that it is now only a calculational matter to algorithmically construct new dipole rings in $D=5$ Einstein-Maxwell-dilaton, including extra rotations and charges. Note that a different systematic construction of this type of dipole black rings was earlier constructed in \cite{Yazadjiev:2006ew}, although their methods do not allow for a straightforward generalisation to include more charges, as the construction \cite{Rocha:2011vv}  does.

In this paper we will construct a new asymptotically flat black ring solution of $D=5$ Einstein-Maxwell-dilaton theory (with KK coupling) featuring rotation, $J_\psi$, $J_\phi$, in two orthogonal planes and both (magnetic) dipole charge $q$ and electric charge $Q$. The full unbalanced solution depends on five parameters (related to the previous four charges, plus the mass $M$), which are subject to one constraint once the balance condition that ensures absence of conical singularities is imposed. Our solution reduces in suitable limits  to the singly spinning dipole ring of  \cite{Emparan:2004wy}, and to the Pomeransky-Sen'kov doubly-spinning neutral ring \cite{Pomeransky:2006bd}. In appendix \ref{app}, we further generalise this solution to include one further parameter.

Dipole charges are peculiar to higher dimensions, and have no counterpart for $D=4$ black holes. For $D=5$ vector-coupled black rings, (magnetic) dipoles arise from integration of the magnetic components of the two-form field strength over a sphere containing the $S^2$ of the ring's $S^1\times S^2$ horizon. Dipoles  make strikingly manifest a continuous black hole non-uniqueness and, in spite of being non-conserved, they nevertheless show up in the first law \cite{Emparan:2004wy,Copsey:2005se}. Following the original construction of the singly spinning dipole black ring in $D=5$ Einstein-Maxwell-dilaton \cite{Emparan:2004wy}, other dipole solutions have been constructed. In \cite{Elvang:2004xi}, black rings of $D=5$ minimal supergravity carrying charges $(M,J_\psi, J_\phi,Q, q)$ were constructed out of more general rings in $U(1)^3$ supergravity ($D=5$ $N=2$ plus two vector multiplets). Regularity of their solutions imposed two constraints among the charges, leaving only three of them independent. 

The conjecture was put forward in \cite{Elvang:2004xi} that a (non-supersymmetric) black ring should exist in minimal supergravity with all those five charges  $(M,J_\psi, J_\phi, Q, q)$ independent. The same conjecture should hold for $D=5$ Einstein-Maxwell-dilaton, a venue very similar to minimal supergravity, especially in the case of KK dilaton coupling. Indeed, both theories share the same U-duality isotropy subgroup $SO(4)$ (or non-compact versions thereof) that preserves given asymptotic conditions. Even if the full duality groups themselves, $G_{2(2)}$ and $SL(4,\mathbb{R})$, are quite different, both theories unify simply upon addition of $n \geq 1$ $N=2$ vector multiplets (see section \ref{sec:setup}), where an overarching $SO(4,n+2)$ duality arises. Finally, any scalar charge carried by the dilaton should be secondary and, thus, always absent from the first law and always a function of the same physical charges $(M,J_\psi, J_\phi, Q, q)$  of minimal supergravity black rings. The new four-independent-parameter solution that we analyse in detail in this paper  gets close to the expected five-parametric solution, in the special case of KK dilaton coupling. Further, preliminary analysis indicates that the solution that we present in the appendix is in fact the most general solution. In any case, these solutions constitute the most general black rings known to date in $D=5$ Einstein-Maxwell-dilaton.

The organisation of the paper is as follows. In order to fix our notation, we briefly review in section \ref{sec:setup} the framework -- the five-dimensional Einstein Maxwell-dilaton-theory -- in which we will work. Section \ref{sec:IMS} provides a detailed account of our six-dimensional inverse-scattering construction leading to a new $D=6$ Ricci flat metric, presented in section \ref{sec:DSCBR}. The full $D=5$ solution is also presented in that section. In section \ref{sec:BRanalysis} we extensively analyse our new $D=5$ black ring, discuss the regularity conditions and ensure its correct flat asymptotics. The computation of the physical parameters is dealt with in section \ref{sec:PhysParam}. In that section we also study some limits of our solution and its charges, both to previously known ones, as a crosscheck, and to extremality. We find that our solution does possess a regular zero temperature limit, where the horizon area and the physical charges remain finite. In section \ref{sec:conclusions} we offer our conclusions and, finally, in appendix \ref{app} we present a more general solution.

\vskip 10pt

\noindent {\it Note added.} As this paper was being prepared for submission, we became aware of~\cite{Chen:2012kd}, where a $D=5$ doubly spinning dipole black ring is also constructed building on~\cite{Rocha:2011vv}. Unlike ours, their solution does not support electric charge, and therefore probes a different slice of the five-dimensional $(M,J_\psi, J_\phi,Q, q)$ parameter space. This seems to be due to a different choice of solitonic transformations in their implementation of the ISM. In \cite{Feldman:2012vd}, a five-parameter black ring is constructed also using the techniques of \cite{Rocha:2011vv}. This is presumably equivalent to the solution that we present in our appendix \ref{app}.

\section{The setup} \label{sec:setup}

The five-dimensional theory we will be focusing on is Einstein-Maxwell-dilaton theory,
\be
S = \frac{1}{16 \pi G_5} \int d^5 x \sqrt{-g} \left(R - \frac{1}{2} \partial_\mu \phi\, \partial^{\mu}\phi - \frac{1}{4} e^{-a \phi}F_{\mu \nu} F^{\mu \nu} \right),
\label{action}
\ee
with the specific dilaton coupling 
\be
a = \frac{2 \sqrt{2}}{\sqrt{3}}.
\label{KKcoupling}
\ee
Only for this coupling does the theory (\ref{action}) arise from circle reduction of six-dimensional vacuum gravity,
\be
S = \frac{1}{16 \pi G_6} \int d^6 x\,\, \sqrt{-g} \,\, R ,
\label{6daction}
\ee
by means of the standard Kaluza-Klein ansatz
\be
ds^2_{6} = e^{\frac{\phi}{\sqrt{6}}} \,ds^2_5 + e^{-\frac{\sqrt{3}\phi}{\sqrt{2}}} (dw + A)^2.
\label{metric6d}
\ee
Here, the metric $ds^2_5$, the dilatonic scalar $\phi$, and the one-form $A$ all take values on five-dimensional spacetime and constitute the fields of the $D=5$ theory (\ref{action}). The internal Kaluza-Klein circle is parametrised by $w$, and we denote by $F=dA$ the abelian two-form field strength of $A$. The $D=5$ fields are subject to the equations of motion
\begin{eqnarray}
&& \nabla_\mu \nabla^\mu \phi +\frac{a}{4}  e^{-a\phi}  F_{\mu \nu} F^{\mu \nu} = 0 \; , \qquad 
\nabla^\mu \big( e^{-a\phi} F_{\mu \nu} \big) = 0 , \nonumber \\
&& R_{\mu \nu} = \frac{1}{2} \partial_\mu \phi  \partial_\nu \phi + \frac{1}{2} e^{-a\phi} \Big( F_{\mu \lambda}  F_\nu{}^\lambda  -\frac{1}{6} g_{\mu \nu}  F_{\lambda \rho} F^{\lambda \rho} \Big) \ ,
\end{eqnarray}
following from the action (\ref{action}).

The smaller supergravity theory that contains (\ref{action}), (\ref{KKcoupling}) is $D=5$ $N=2$ supergravity coupled to a vector multiplet. The bosonic content of this theory includes the metric and one vector in the supergravity multiplet, plus another vector and the dilaton in the vector multiplet. The resulting Chern-Simons coupling $C_{IJK}$, $I,J,K=0,1$, is such that two different  consistent subtruncations are allowed: to minimal $N=2$ supergravity and to Einstein-Maxwell-dilaton (\ref{action}). To see how that field content arises, simply enlarge the $D=6$ GR action (\ref{6daction}) to that of  minimal, $D=6$ $(1,0)$ supergravity by means of a two-form $B^+$ with self-dual three-form field strength. Due to its self-duality, $B^+$ gives rise to only one vector upon circle reduction, thus indeed yielding supergravity plus one vector multiplet. Note that $D=6$ $(1,0)$ supergravity itself arises from heterotic on K3 or, equivalently, from the NS sector of type II on K3, or $T^4$, followed by further truncation.

Black rings with a single rotation along the $S^1$ and with dipole charge but no electric charge were first constructed \cite{Emparan:2004wy} in the theory (\ref{action}) with arbitrary dilation coupling $a$ by educated guess-work. In~\cite{Rocha:2011vv}, two of us showed how to systematically construct the solutions of  \cite{Emparan:2004wy},  in the specific case of Kaluza-Klein dilation coupling  (\ref{KKcoupling}), through the ISM in six dimensions. We will now show how the analysis of \cite{Rocha:2011vv} can be extended to generate both electric charge and a second rotation along the $S^2$.

\section{Inverse scattering construction of the solution} \label{sec:IMS}

\begin{figure}[t!]
\centering{
\includegraphics[width=140mm,height=40mm]{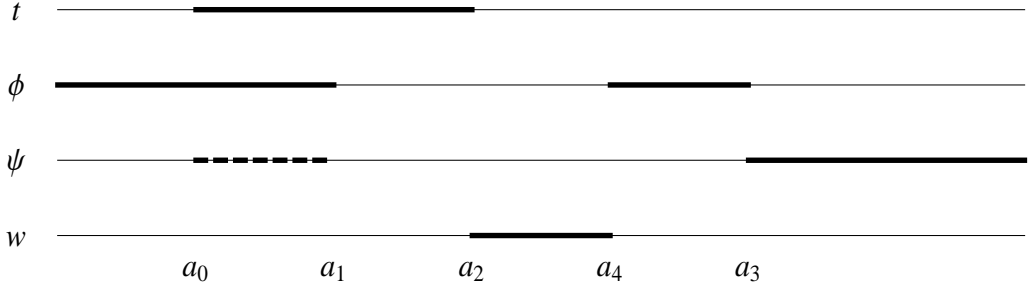}}
\caption{The figure depicts sources for the seed metric $G_0$.  The rods
  are located in $(\rho, z)$ space at the $z$-axis with $\rho=0$. The solid rods have
  positive density and the dashed rod has negative density. They add up to an infinite
  rod with uniform density such that $\det G_0 = -\rho^2$.}
\label{roddiagram}
\end{figure}

Our starting point is the rod configuration shown in figure~\ref{roddiagram}, corresponding to the seed solution of the six-dimensional uplift of the dipole black ring. The thick solid lines correspond to rod sources of uniform density +1/2, and the dashed line corresponds to a rod source of uniform density $-$1/2. The rod in the $t$ direction corresponds to the horizon. In figure~\ref{roddiagram}, when $a_0 = a_1$ and $a_4 = a_2$, the negative density rod and the rod in the $w$ direction disappear and the solution describes a neutral static black ring times a flat direction $w$. This is an unbalanced configuration. The negative density rod is included in the seed following~\cite{Elvang:2007rd} to facilitate adding the $S^1$ angular momentum to the ring. The positive density rod $[a_2,a_4]$ along the $w$ direction is included to facilitate adding dipole charge to the ring.

The seed metric corresponding to the rod configuration of figure~\ref{roddiagram} is given by
\be
ds^2_6 = (G_0)_{ab}\; dx^a dx^b+ e^{2\nu_0}(d\rho^2+dz^2),
\label{confmetric}
\ee
where
\be
G_0 = \verb+diag+\left\{ -\frac{\mu_0}{\mu_2}, \frac{\rho^2 \mu_4}{\mu_1 \mu_3}, \frac{\mu_1 \mu_3}{\mu_0}, \frac{\mu_2}{\mu_4} \right\}, \qquad \det G_0 = - \rho^2.
\label{seed}
\ee
and the conformal factor of the seed is
\be
e^{2\nu_0} = k^2 \frac{\mu_1 \mu_3}{\mu_0} \frac{(\mu_0 \mu_1+\rho^2)(\mu_0 \mu_2+\rho^2)(\mu_0 \mu_3+\rho^2)(\mu_1 \mu_4+\rho^2)(\mu_2 \mu_4+\rho^2)(\mu_3 \mu_4+\rho^2)}{(\mu_1 \mu_3+\rho^2)^2 \prod_{i=0}^{4}(\mu_i^2+\rho^2)}.
\label{e2nu}
\ee
The integration constant $k$ will be fixed in the next subsection. Our ordering of coordinates is $ x^a=(t, \phi, \psi, w)$, with $t$ corresponding to the timelike coordinate, $\phi$ describing the azimuthal angle on the $S^2$ and $\psi$ being the angle along the $S^1$ component of the ring. The $\mu_i$'s are pole trajectories of the dressing matrices, commonly referred to as solitons, $\mu_i = \sqrt{\rho^2 + (z-a_i)^2} - (z-a_i)$, and $a_i$, $i=0,1,\ldots, 4$, are the rod endpoints. In writing these expressions we have followed established conventions from the literature.  We refer the reader to e.g. \cite{Emparan:2008eg} for notational details.

We assume the ordering
\be
a_0 \le a_1 \le a_2 \le a_4 \le a_3. \label{ordering}
\ee
The labeling (and hence the ordering) is a little unusual, but is motivated to simplify the presentation of the solution after the inverse scattering transformation. We will see that the endpoints $a_0$ and $a_4$ will not appear in the rod diagram of the final solution.

The seed solution \eqref{confmetric}--\eqref{e2nu} with ordering \eqref{ordering} is singular and not  of direct physical interest itself. However, applying the ISM we will be able to find a meaningful new solution by dressing this seed metric, that is, by removing and adding solitons. Details on the ISM and rod diagram representations can be found in ~\cite{BZ, Emparan:2001wk, Harmark:2004rm, Kleihaus:2010pr}.

In more detail, these are the steps that we follow in order to generate the (six-dimensional uplift of the) doubly spinning dipole ring solution by a 4-soliton transformation:
\begin{enumerate}
\item Perform the following four transformations on the seed solution \eqref{seed}:
$(i)$ remove a soliton at $z=a_0$ with trivial Belinski-Zakharov (BZ) vector $(0,0,1,0)$,
$(ii)$ remove a soliton at $z=a_1$ with trivial BZ vector $(0,1,0,0)$,
$(iii)$ remove an anti-soliton at $z=a_2$ with trivial BZ vector $(0,1,0,0)$,
$(iv)$ remove a soliton at $z=a_4$ with trivial BZ vector $(0,0,0,1)$,
and supplement all this with a rescaling of the metric by an overall factor $\zeta=\rho^2/(\mu_1 \mu_4)$.
The resulting metric is
\be
G_0' = \verb+diag+\left\{- \frac{\mu_0\,\rho^2}{\mu_1\,\mu_2\,\mu_4}, \frac{\rho^4}{\mu_2^2\,\mu_3},- \frac{\mu_0\, \mu_3}{\mu_4},- \frac{{\mu}_2}{\mu_1} \right\}.
\label{seed1}
\ee
This metric will be the seed for our next soliton transformation.
\item The generating matrix corresponding to \eqref{seed1}, in a form which is convenient for the solitonic transformation that will follow, reads
\be
\psi_0' = \verb+diag+\left\{ \frac{(\bar{\mu}_1-\lambda)\,(\bar{\mu}_4-\lambda)}{(\bar{\mu}_0-\lambda)\,(\mu_2-\lambda)}, \frac{(\mu_3-\lambda)\,(\bar{\mu}_3-\lambda)^2}{(\mu_2-\lambda)^2},\frac{-(\mu_3-\lambda)\, (\bar{\mu}_4-\lambda)}{(\bar{\mu}_0-\lambda)},\frac{-(\bar{\mu}_1-\lambda)\,({\mu}_2-\lambda)}{(\mu_3-\lambda) \,(\bar{\mu}_3-\lambda)} \right\}.
\label{Gseed1}
\ee
\item Perform now a 4-soliton transformation with seed $G_0'$ and undo the rescaling. More precisely 
$(i)$ add a soliton at $z=a_0$ with BZ vector $(c_1,0,1,0)$,
$(ii)$ add a soliton at $z=a_1$ with BZ vector $(0,1,b_1,0)$,
$(iii)$ add an anti-soliton at $z=a_2$ with BZ vector $(b_2,1,0,b_3)$,
$(iv)$ add a soliton at $z=a_4$ with BZ vector $(0,c_2,0,1)$,
 and rescale by dividing  by $\zeta$. Denote the final metric by $G$. The final rescaling ensures that $\det G = - \rho^2$.
\item Construct the conformal factor. A straightforward computation gives the conformal factor of the new metic as
$e^{2\nu}=e^{2\nu_0} \frac{\det\Gamma}{\det\Gamma_0}$ where the matrix $\Gamma$ effectively implements the addition of solitons~\cite{Emparan:2008eg, Iguchi:2011qi} and  $\Gamma_0=\Gamma_{c_1,b_1,b_2,b_3,c_2=0}$.
\end{enumerate}
The result $(G,e^{2\nu})$ is the six-dimensional solution we want. In the next section we present the resulting solution in Weyl coordinates. The analysis of  regularity and asymptotic flatness in $D=5$ is subsequently performed in section \ref{sec:BRanalysis}. For simplicity, in this paper we will analyse in detail the case for which $b_3 = 0$, which we set henceforth. We will see that this leads to a four-parameter solution once the regularity and asymptotic conditions are imposed. The most general solution with non-vanishing $b_3$ can be found in Appendix \ref{app}.

\section{The electrically charged doubly spinning dipole black ring solution} \label{sec:DSCBR}
The new six-dimensional Ricci-flat metric is 
\be
ds^2_6 = (G)_{ab}\; dx^a dx^b+ e^{2\nu}(d\rho^2+dz^2),\qquad x^a=(t, \phi, \psi, w)\,,
\label{newmetric}
\ee
where
\be
G = (G'_0-N_v^T\,\Gamma^{-1}\,N_v)\,\zeta^{-1}, \qquad e^{2\nu}=e^{2\nu_0} \frac{\det\Gamma}{\det\Gamma_0}\,.
\label{newGconf}
\ee
Here,  $(G'_0,e^{2\nu_0})$ are the seed metric expressions given in \eqref{seed1} and \eqref{e2nu}, and $\zeta^{-1}=\mu_1 \mu_4/\rho^2$ has also been defined previously. The determinant of $\Gamma_0$ is
\be
\det\Gamma_0=-\frac{ \mu_1 \mu_2 \mu_3^5 \mu_4 (\mu_1-\mu_2)^2 (\mu_3-\mu_4)^2 Z_{00} Z_{12}^2 Z_{22}^3 Z_{34}^2}{\rho^4 \mu_0  (\mu_0-\mu_3)^2 (\mu_1-\mu_3)^2 (\mu_2-\mu_3)^4 (\mu_2-\mu_4)^2 Z_{04}^2 Z_{11} Z_{13}^4 Z_{14}^2 Z_{23}^2 Z_{44} }\,,
\ee
where we have introduced the function $Z_{ij}=\rho^2+\mu_i\mu_j$. The remaining quantities that appear in (\ref{newGconf}) are the matrices $N_v$ and $\Gamma$. Their explicit $(\rho,z)$-dependent expressions are 
\be 
N_v=
 \begin{pmatrix}
   \frac{c_1 \rho^2 \left(\rho^2+\mu_0^2\right) (\mu_2-\mu_0)}{\mu_0 \left(\rho^2+\mu_0 \mu_1\right) \mu_2 \left(\rho^2+\mu_0 \mu_4\right)} & 0& \frac{\left(\rho^2+\mu_0^2\right) \mu_3}{\mu_0 (\mu_3-\mu_0) \left(\rho^2+\mu_0 \mu_4\right)} & 0\\
   0 & \frac{\rho^4 (\mu_1-\mu_2)^2 \mu_3}{\mu_1 \mu_2^2 (\mu_3-\mu_1) \left(\rho^2+\mu_1 \mu_3\right)^2} & \frac{b_1 \left(\rho^2+\mu_0 \mu_1\right) \mu_3}{\mu_1 (\mu_3-\mu_1) \left(\rho^2+\mu_1 \mu_4\right)} &0\\
    \frac{b_2 (\mu_0-\mu_2) \left(\rho^2+\mu_2^2\right)}{\rho^2 (\mu_2-\mu_1) (\mu_2-\mu_4)} & -\frac{\left(\rho^2+\mu_2^2\right)^2 \mu_3}{\rho^2 (\mu_2-\mu_3)^2 \left(\rho^2+\mu_2 \mu_3\right)} & 0 & 0\\
   0 & \frac{c_2 \rho^4 \mu_3 (\mu_2-\mu_4)^2}{\mu_2^2 (\mu_3-\mu_4) \mu_4 \left(\rho^2+\mu_3 \mu_4\right)^2} & 0 &\frac{\mu_2 (\mu_3-\mu_4) \left(\rho^2+\mu_3 \mu_4\right)}{\mu_3 (\mu_2-\mu_4) \mu_4 \left(\rho^2+\mu_1 \mu_4\right)}\\
 \end{pmatrix}
\ee
and
\be
\Gamma=
 \begin{pmatrix}
   \Gamma_{11} & \Gamma_{12}& \Gamma_{13} & 0\\
   \Gamma_{12} & \Gamma_{22} & \Gamma_{23} &\Gamma_{24}\\
   \Gamma_{13} & \Gamma_{23} & \Gamma_{33} &\Gamma_{34}\\ 
   0 & \Gamma_{24} & \Gamma_{34} &\Gamma_{44}\\
 \end{pmatrix}\,,
\ee
with components
\be \label{GammaComps}
\Gamma_{11} = \frac{-c_1^2 \rho^2 \mu_1 \mu_4 (\mu_0-\mu_2)^2 (\mu_0-\mu_3)^2 Z_{00} - \mu_2 \mu_3 \mu_4 Z_{00} Z_{01}^2}{\mu_0 \mu_2 (\mu_0-\mu_3)^2 Z_{01}^2 Z_{04}^2}\nonumber\,,\\
\ee
\be
\Gamma_{12} = \frac{b_1 \mu_3 \mu_4 Z_{00} }{\mu_0 (\mu_0-\mu_3) (\mu_3-\mu_1) Z_{04}Z_{14}}\nonumber\,,\qquad \Gamma_{13}=\frac{b_2 c_1 \mu_1 \mu_4 (\mu_0-\mu_2) Z_{00} Z_{22} }{\rho^2 \mu_0 (\mu_2-\mu_1) (\mu_2-\mu_4)Z_{01} Z_{04}}\nonumber\,,
\ee
\be
\Gamma_{22}= \frac{-b_1^2  \mu_2^2 \mu_3  \mu_4 Z_{01}^2 Z_{13}^4+\rho^4 \mu_0 \mu_3^3 (\mu_1-\mu_2)^4  Z_{14}^2}{\mu_0 \mu_2^2 (\mu_1-\mu_3)^2 Z_{11} Z_{13}^4 Z_{14}^2}\,, \nonumber
\ee
\be
\Gamma_{23}= \frac{ \mu_3^3 (\mu_1-\mu_2) Z_{22}^2}{\rho^2 (\mu_1-\mu_3) (\mu_2-\mu_3)^2 Z_{13}^2 Z_{23}}\,,\qquad \Gamma_{24}= -\frac{c_2 \rho^4 \mu_3^3 (\mu_1-\mu_2)^2(\mu_2-\mu_4)^2}{\mu_2^2 (\mu_1-\mu_3) (\mu_3-\mu_4) Z_{13}^2 Z_{14} Z_{34}^2}\,,
\ee
\be
\Gamma_{33}= \frac{\mu_2 Z_{22} \left(\mu_0 \mu_2 \mu_3^3 (\mu_1-\mu_2)^2 (\mu_2-\mu_4)^2 Z_{22}^2 - b_2^2 \rho^2 \mu_1 \mu_4 (\mu_0-\mu_2)^2 (\mu_2-\mu_3)^4 Z_{23}^2 \right)}{\rho^6 \mu_0 (\mu_1-\mu_2)^2 (\mu_2-\mu_3)^4 (\mu_2-\mu_4)^2 Z_{23}^2}\,, \nonumber
\ee
\be
\Gamma_{34}= \frac{c_2 \mu_3^3 (\mu_2-\mu_4) Z_{22}^2 }{\rho^2 (\mu_2-\mu_3)^2(\mu_3-\mu_4) Z_{23} Z_{34}^2}\,, \qquad \Gamma_{44}= \frac{c_2^2 \rho^4 \mu_3^5 (\mu_2-\mu_4)^6 Z_{14}^2-\mu_1 \mu_2^3 (\mu_3-\mu_4)^4 Z_{34}^6}{\mu_2^2 \mu_3^2 (\mu_2-\mu_4)^2 (\mu_3-\mu_4)^2  Z_{14}^2 Z_{34}^4 Z_{44}}\,.\nonumber
\ee

There still remains a final step. After applying the solitonic transformations, the new six-dimensional Ricci-flat solution (\ref{newmetric})--(\ref{GammaComps}) has a non standard alignment of the rods. In order to enforce their correct physical alignment and, as we will discuss shortly, be able to explicitly show the desired asymptotic behaviour, we now perform a linear transformation of the $x^a$ coordinates by means of a matrix
\be
\label{AmatTrans}
A=
 \begin{pmatrix}
 l_0 & q_0& p_0 & s_0\\
 0 & q_1 & p_1 & s_1\\
 0 & q_2 & p_2 & s_2\\ 
 l_3& q_3 & p_3 & s_3\\
 \end{pmatrix}\ ,
\ee
under which $G$ in (\ref{newmetric}) transforms into
\be\label{metrictransf}
\tilde{G}=A^T\, G\,A \,. 
\ee
Although we take $A$ to be general for now, we will restrict it to lie in  $SL(4,\mathbb{R})$ in the next section. The final, Ricci-flat six-dimensional metric  is (\ref{newmetric})--(\ref{GammaComps}), with $G$ replaced by $\tilde{G}$. 

Let us now tally up the number of parameters contained in the raw $D=6$ Ricci-flat metric thus obtained: as it stands, the metric  (\ref{newmetric})-(\ref{metrictransf}) contains 24 parameters. These can be classified into three groups: $(1)$ the rod endpoints ($a_0,a_1,a_2,a_3,a_4$), making up for 5 parameters; $(2)$ the 4 BZ-parameters ($c_1, c_2, b_1, b_2$); and $(3)$ 15 more parameters including the integration contant $k$ arising in the conformal factor $e^{2\nu}$, together with the alignment transformation parameters ($l_0, l_3$, $q_i,p_i,s_i)$, $i=0, \ldots ,3$, entering the linear transformation matrix $A$ in (\ref{AmatTrans}). Strictly speaking, the latter 14 parameters merely correspond to a coordinate transformation of the metric whose only role is to make manifest the desired asymptotic behaviour. Following standard practice, we nevertheless include them in the count. 

We can now write down the corresponding $D=5$ Einstein-Maxwell-dilaton 24-parameter solution. Working out the lifting formula (\ref{metric6d}) backwards, we can write the $D=5$ metric, gauge field $A=A_i dx^i$, and dilaton $\phi$ in terms of the $D=6$ metric components as
\be\label{5dsol}
ds_5^2=G^{(5)}_{ij} dx^i dx^j+ \tilde{G}^{1/3}_{ww} \,e^{2\nu}(d\rho^2+dz^2) \,,\qquad \text{where} \qquad G^{(5)}_{ij}=\left(\tilde{G}_{ij}-\frac{\tilde{G}_{iw}\tilde{G}_{jw}}{\tilde{G}_{ww}}\right)\tilde{G}^{1/3}_{ww}\,,
\ee
\be\label{5dsol2}
A_i=\tilde{G}_{iw}/\tilde{G}_{ww}\,,\qquad e^{-\frac{\sqrt{3}}{\sqrt{2}}\phi}=\tilde{G}_{ww} \ ,
\ee
for $x^i=(t,\phi,\psi)$. As we will see in the next section, the 24 parameters of this solution reduce to only 4 independent ones, once the correct regularity and $D=5$ asymptotic conditions are imposed. These four parameters just reflect the five (constrained) physical charges $(M,J_\psi, J_\phi,Q, q)$, to be computed in section \ref{sec:PhysParam}, that the solution (\ref{5dsol}), (\ref{5dsol2}) carries. Finally, by construction, the full unbalanced solution (\ref{5dsol}), (\ref{5dsol2}) reduces both to the unbalanced \cite{Rocha:2011vv}  singly spinning dipole ring \cite{Emparan:2004wy} and to the unbalanced  \cite{Chen:2011jb,RRunp} doubly spinning neutral ring \cite{Pomeransky:2006bd}. All these considerations lead us to the conclusion that the balanced version of (\ref{5dsol}), (\ref{5dsol2}) is indeed a new 4-parameter generalisation of the black rings of \cite{Emparan:2004wy,Pomeransky:2006bd}.

\section{Regularity analysis} \label{sec:BRanalysis}

Having obtained the full $D=6$ and $D=5$ solutions, we now proceed with the regularity analysis. We first enforce a correct rod alignment in subsection \ref{subsec:RodStr}, which will allow us to explicitly fix the asymptotics in subsection \ref{subsec:Asympt}. Finally, in \ref{subsec:Reg} we study the absence of conical singularities, closed time-like curves and Dirac-Misner strings. Going through the regularity checklist fixes many of the free parameters. We will track those down at the end of subsection \ref{subsec:Reg}.

As a consistency check, we have verified that all the constraints on the parameters that we derive below reduce in suitable limits to those corresponding to the singly spinning \cite{Rocha:2011vv,Emparan:2004wy} or neutral \cite{Chen:2011jb,RRunp,Pomeransky:2006bd} cases.

\subsection{Rod structure} \label{subsec:RodStr}

\begin{figure}[t!]
\centering{
\includegraphics[width=140mm,height=40mm]{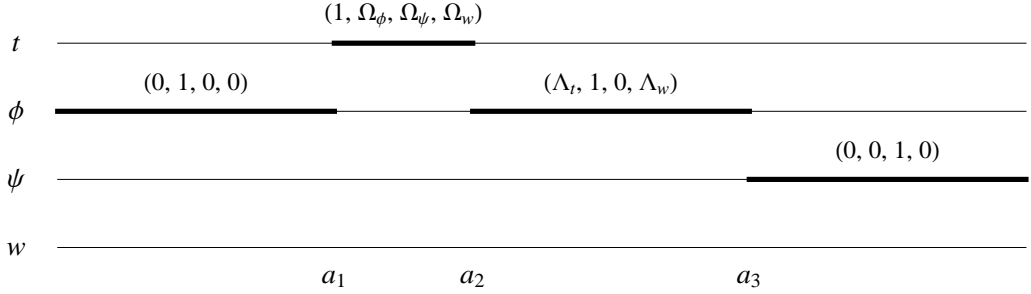}}
\caption{The figure depicts sources for the six-dimensional lift of the doubly spinning charged black ring. The direction of each rod is indicated. The points $a_0$ and $a_4$ are not indicated in the figure, as they no longer represent real turning points.}
\label{roddiagram3}
\end{figure}

The rod diagram in figure~\ref{roddiagram3} shows the standard alignment of the rods so that the metric takes a simple diagonal form at infinity. Imposing this asymptotic behaviour on the new solution fixes some of its parameters. We find:

\begin{itemize}
\item The semi-infinite rod $z \in (-\infty, a_0]$ has direction $(0,1,0,0)$. This can be achieved by setting
\bea
q_0&=&\frac{2 (a_0-a_2) (a_1-a_2) (a_0-a_3) (a_2-a_4) b_1 c_1\, q_1}{(a_1-a_2)^2 (a_2-a_4)+2 (a_0-a_2) (a_0-a_3) (a_2-a_3)^2 b_1 b_2 c_1}\,, \\
q_2&=&\frac{(a_0-a_1) (a_1-a_2) (a_2-a_4) b_1\, q_1}{(a_1-a_2)^2 (a_2-a_4)+2 (a_0-a_2) (a_0-a_3) (a_2-a_3)^2 b_1 b_2 c_1}\,, \\
q_3&=&0\,.
\eea
The mixing coefficient $q_1$ can be fixed by requiring the rod direction to be normalized, but for now we leave it unconstrained.

\item Similarly, the rod $z \in [a_0, a_1]$ has direction $(0,1,0,0)$. This occurs if and only if
\be
|c_1|= \sqrt{\frac{(a_1-a_0)}{2 (a_2-a_0)(a_3-a_0)}}\,.
\label{constraint:c1}
\ee

\item The semi-infinite rod $z \in [a_3, \infty)$ has rod direction $(0,0,1,0)$. This can be achieved by setting
\bea
p_0&=&-\frac{2 (a_0-a_1) (a_0-a_2) (a_2-a_3)^2 b_1 b_2\,p_2}{(a_1-a_2)^2 (a_2-a_4)+2 (a_0-a_2) (a_0-a_3) (a_2-a_3)^2 b_1 b_2 c_1}\,, \\
p_1&=&\frac{(a_0-a_1) (a_1-a_2) (a_2-a_4) b_1\,p_2}{(a_1-a_2)^2 (a_2-a_4)+2 (a_0-a_2) (a_0-a_3) (a_2-a_3)^2 b_1 b_2 c_1}\,, \\
p_3&=& - \frac{(a_0-a_1) (a_1-a_2) (a_2-a_4)^3 b_1 c_2\,p_2}{(a_4-a_3)^2 \left((a_1-a_2)^2 (a_2-a_4)+2 (a_0-a_2) (a_0-a_3) (a_2-a_3)^2 b_1 b_2 c_1 \right)}\,.\label{p3}
\eea
The mixing coefficient $p_2$ can be fixed by requiring the rod direction to be normalized but, again, we leave it unconstrained for now.

\item The finite rods $[a_2,a_4]$ and $[a_4,a_3]$ are aligned\footnote{Note that this condition can be relaxed so that in general they do not have the same normalized direction.} with direction $(\Lambda_t,1,0, \Lambda_w)$. We therefore require that these rods be parallel, which fixes
\be
|c_2|=\sqrt{\frac{2 (a_3-a_4)^5}{(a_4-a_1) (a_4-a_2)^3}}\,.
\label{constraint:c2}
\ee
and
\be
2\,b_1 b_2^2 \,d_{10}\,d_{20}\,d_{21} {d_{32}}^3 + b_1 d_{10}\, {d_{21}}^2 {d_{42}}^2
- 2\,b_2 c_1 \, d_{20} \,d_{32}\, d_{42}\left( {d_{21}}^3 + b_1^2 \,d_{10}\, {d_{31}}^2 \right)  = 0\,,
\label{constraint_b1b2}
\ee
where we have introduced the notation $d_{ij} \equiv a_i-a_j$.
This follows from the requirement that the $\phi$- and $\psi$-components of the direction of rods $(-\infty, a_1]$ and $[a_2,a_3]$ coincide, which also fixes
\be \label{s1s2Cond}
s_1= s_2=0\,.
\ee
Once all the constraints are imposed we find that
\be \label{LambdaConds}
\Lambda_t = \frac{2 d_{34}^2 \left( 2 b_1 b_2 c_1 d_{20} d_{31}^2 d_{32} - d_{21}^2 d_{41} \right)q_{1} s_{0}}{c_{2} d_{41} \left( 2 b_1 b_2 c_1 d_{20} d_{31}^2 d_{32} - d_{21}^2 d_{42} \right) (l_{3} s_{0}-l_0 s_3)}\,,\qquad \Lambda_w=-\Lambda_t\frac{l_0}{s_0}\,.
\ee
When no rotation and no electric charge is added to the $S^2$, i.e. $b_1=b_2=0$, $s_0=l_3=0$ and $l_0=s_3=q_1=1$, these reduce to the corresponding values \cite{Rocha:2011vv} $\Lambda_t=0$ and $\Lambda_w=\sqrt{\frac{2 (a_4-a_2) (a_4-a_1) }{a_3-a_4}}$ found for the singly spinning dipole ring of \cite{Emparan:2004wy}.

\item The finite timelike rod $[a_1,a_2]$ has rod direction $(1,\Omega_\phi,\Omega_\psi,\Omega_w)$. 
Without imposing any of the above constraints the components of this vector are 
\bea
\Omega_\phi &=& \frac{d_{21} d_{41} d_{42} \left[ c_1(Q_{31} l_0+Q_{10} l_3) - Q_{21} l_3 \right] + \left(b_1 c_1 d_{31}^2 d_{42} + b_2 d_{32}^2 d_{41} \right)(Q_{32} l_0+Q_{20}l_3)}{D_\Omega},\label{Omega2}\\
\Omega_\psi &=& \frac{d_{21} d_{41} d_{42} \left[ c_1(P_{31} l_0+P_{10} l_3) - P_{21} l_3\right] + \left(b_1 c_1 d_{31}^2 d_{42} + b_2 d_{32}^2 d_{41} \right)(P_{32} l_0+P_{20}l_3)}{D_\Omega}\,,\label{Omega3}\\
\Omega_w &=& \frac{d_{21} d_{41} d_{42} \left[ c_1(S_{31} l_0+S_{10} l_3) - S_{21} l_3 \right] + \left(b_1 c_1 d_{31}^2 d_{42} + b_2 d_{32}^2 d_{41} \right)(S_{32} l_0+S_{20}l_3)}{D_\Omega}\,,\label{Omega4}
\eea
where
\bea
D_\Omega &=& d_{21}d_{41}d_{42} \left[(S_{21} s_3 - S_{31} s_2 + S_{32} s_1) - c_1 (S_{10} s_3- S_{30} s_1 +S_{31} s_0)\right] \nonumber\\
    && - \left( b_1 c_1 d_{31}^2 d_{42} + b_2 d_{32}^2 d_{41} \right)(S_{20} s_3 - S_{30} s_2 + S_{32} s_0) \,.\label{OmegaD}
\eea
To reduce cluttering we have introduced the notation $Q_{ij}=p_i\,s_j-p_j\,s_i$, $\;P_{ij}=s_i\,q_j-s_j\,q_i$ and $S_{ij}=q_i\,p_j-q_j\,p_i$, with $i,j\in\{0,1,2,3\}$.
The expressions for the components $\Omega_i$ reduce somewhat upon imposing the above constraints.
Note that when no rotation is added to the $S^2$ the direction of the timelike rod becomes $(1,0,-c_1,0)$, again in perfect agreement with~\cite{Rocha:2011vv}.
 
\end{itemize}

\bigskip
Following the same strategy as in~\cite{Chen:2011jb}, one can now use the constraint~\eqref{constraint_b1b2} to fix $a_0$. More concretely, the relevant solution for $b_2$ is
\be
b_2=\frac{(a_2-a_1)^2(a_4-a_2)}{\sqrt{2(a_1-a_0)(a_2-a_0)(a_3-a_0)} (a_3-a_2)^2}  \frac{\varpi-\sqrt{\varpi^2-4\vartheta}}{2b_1}\,,
\ee
where
\be
\varpi= 1 + \frac{(a_1-a_0)(a_3-a_1)^2}{(a_2-a_1)^3} b_1^2\,, \qquad \qquad  \vartheta= \frac{(a_1-a_0)(a_3-a_0)(a_3-a_2)}{(a_2-a_1)^3} b_1^2\,.
\label{def:varpi}
\ee
Further defining
\be
\alpha= \frac{\varpi+\sqrt{\varpi^2-4\vartheta}}{2}\,, \qquad \qquad  \beta= \frac{\varpi-\sqrt{\varpi^2-4\vartheta}}{2}\,,
\label{def:alpha}
\ee
we may express the BZ parameters $b_1$ and $b_2$ as
\be
b_1= \sqrt{\frac{(a_2-a_1)^3 \alpha \beta}{(a_1-a_0)(a_3-a_0)(a_3-a_2)}}\,, \qquad \qquad  b_2= \sqrt{\frac{(a_2-a_1)(a_4-a_2)^2 \beta}{2(a_2-a_0)(a_3-a_2)^3 \alpha}}\,.
\label{constraint:b1b2}
\ee
In addition, the relations~\eqref{def:varpi} and~\eqref{def:alpha} imply
\be
a_0 = a_3 - \frac{(a_3-a_1)^2 \alpha \beta}{(a_3-a_2)(\alpha+\beta-1)}\,.
\label{constraint:a0}
\ee
This condition is valid whenever $\alpha\neq 1$ and $\beta\neq 0$. With this parametrization, having no rotation on the $S^2$ corresponds to first taking the limit $\alpha \rightarrow 1$ and then $\beta \rightarrow 0$, in which case the non-trivial components of the rod directions $\Omega_w, \Omega_\phi, \Omega_\psi, \Lambda_{w}$ reduce to the expressions given previously in \cite{Rocha:2011vv}. Note that when $b_1=b_2=0$ the constraint (\ref{constraint:a0}) does not arise given that~\eqref{constraint_b1b2} is automatically satisfied.

\subsection{Asymptotics} \label{subsec:Asympt}
\label{section:sixfived}

We can now verify that the rod alignment discussed in subsection \ref{subsec:RodStr} is compatible with explicit $D=5$ flat asymptotics. These become manifest upon fixing a few more parameters. The $D=6$ metric (\ref{newmetric})-(\ref{metrictransf}) must have KK asymptotics, $\mathbb{R}^5 \times S^1$,
\be\label{AFmetric6}
ds^2_6 \rightarrow -dt^2+dr^2+ r^2 (d\theta^2+\sin^2 \theta \,d\psi^2+\cos^2 \theta \,d\phi^2) + dw^2 \, ,
\ee
so that its $D=5$ counterpart (\ref{5dsol}) is explicitly asymptotically flat,
\be\label{AFmetric5}
ds^2_5 \rightarrow -dt^2+dr^2+ r^2 (d\theta^2+\sin^2 \theta \,d\psi^2+\cos^2 \theta \,d\phi^2)  \,.
\ee
Here, for the sake of simplicity we have relabeled the coordinates, $\tilde{x}^a\rightarrow x^a$.

In order to check the asymptotics of our solution, we introduce asymptotic coordinates $(r,\theta)$ as
\be\label{coordchange}
\rho=\frac{1}{2}\, r^2 \sin 2 \theta \,,\qquad z=\frac{1}{2}\, r^2 \cos 2 \theta\,,
\ee
and impose that the $D=5$ metric (\ref{5dsol}) approaches (\ref{AFmetric5}) as $r\rightarrow \infty$. This requirement leads to $s_1=s_2=0$ (a condition also imposed by rod alignment, see equation (\ref{s1s2Cond}) above) along with
\be \label{conkConf}
k^2=\frac{d_{21}^4\, d_{42}^2}{(2\, b_1 b_2 c_1\, d_{20} \,d_{30}\, d_{32}^2 - d_{21}^2 \,d_{42})^2-b_1^2\,d_{10}^2\, d_{21}^2d_{42}^2}\,.
\ee
Already using these conditions in order to simplify the presentation, we find the following fall-off of the $D=5$ metric components:
\be
G^{(5)}_{tt}\rightarrow -\frac{(l_3 \,s_0-l_0\,s_3)^2}{(s_3^2-s_0^2)^{2/3}}+ O[1/r^2]\,,
\ee
\be
 G^{(5)}_{\psi\psi}\rightarrow \frac{d_{21}^4 d_{42}^2 \,p_2^2 \,(s_3^2-s_0^2)^{1/3} }{k^2\,(2\, b_1 b_2 c_1\, d_{20} \,d_{30} \,d_{32}^2 - d_{21}^2 d_{42})^2}\,r^2\sin^2\theta + O[r^0]
\ee
\be
G^{(5)}_{\phi\phi}\rightarrow  \frac{d_{21}^4 d_{42}^2 \,q_1^2\, (s_3^2-s_0^2)^{1/3} }{k^2\,(2\, b_1 b_2 c_1 \,d_{20} \,d_{30} \,d_{32}^2 - d_{21}^2 d_{42})^2}\,\,r^2\cos^2\theta  + O[r^0] 
\ee
\be
 G^{(5)}_{\phi\psi}\rightarrow 0+ O[1/r^2] \,,\,\,\,\, G^{(5)}_{t\phi}\rightarrow  0+ O[1/r^2] \,,\,\,\,\, G^{(5)}_{t\psi}\rightarrow 0+ O[1/r^2] 
\ee
\be
\tilde{G}^{1/3}_{ww} e^{2\nu}(d\rho^2+dz^2) \rightarrow (dr^2+r^2 d\theta^2) (1+O[1/r^2])
\ee
where we have employed the coordinates (\ref{coordchange}). Comparing with (\ref{AFmetric5}) we see that we must further impose
\be\label{conq1p2}
q_1=p_2=\frac{k\,(2\, b_1 b_2 c_1\, d_{20} \,d_{30} \,d_{32}^2 - d_{21}^2 d_{42})}{d_{21}^2 d_{42}  }\,,
\ee
with $k$ now given by (\ref{conkConf}), and a further restriction in the $(l_0, l_3, s_0, s_3)$ sector. Before specifying the latter, we turn to the fall-off of the $D=5$ gauge field (\ref{5dsol2}),
\be\label{gauges}
A_t\rightarrow -\frac{(l_0\,s_0-l_3\, s_3)}{(s_3^2-s_0^2)}+ O[1/r^2]\,,\,\,\,\, A_{\phi}\rightarrow 0+O[1/r^2]\,,\,\,\,\, A_\psi\rightarrow 0+ O[1/r^2] .
\ee
Imposing the unit determinant condition on the matrix $A$ in (\ref{AmatTrans}), the asymptotic consistency condition $G^{(5)}_{tt}\rightarrow -1+ O[1/r^2]$ and the gauge choice $A_t\rightarrow 0+ O[1/r^2]$, only one (boost) parameter $\sigma$ survives in the $(l_0, l_3, s_0, s_3)$ sector:
\be \label{l0l3mat}
\begin{pmatrix}
 l_0& s_0\\
  l_3 & s_3 \\
\end{pmatrix}
=\begin{pmatrix}
   \cosh\sigma& \sinh\sigma\\
  \sinh\sigma & \cosh\sigma \\
\end{pmatrix}
\, .
\ee
Finally, with all these conditions, we find that the $D=5$ dilaton (\ref{5dsol2}) also acquires the correct asymptotics:
\be
\phi \rightarrow 0+O[1/r^2] .
\ee

Note that the values of $b_1$ and $b_2$ are not fixed by the $D=5$ asymptotically flat boundary conditions. This implies that the inner rod can have an arbitrary direction. Also, the parameter $\sigma$ in (\ref{l0l3mat}) or, equivalently $s_0$, will become trivial when we later require absence of Dirac-Misner strings.

\subsection{Regularity and final parameter counting} \label{subsec:Reg}

We now run the usual regularity tests. We first check for further restrictions on the parameters that render the solution conical-singularity-free, and then turn to check the absence of other possible pathologies.

\subsubsection*{Conical singularities}

Conical singularities would arise if the angular variables could not be given correct periods. We will now show that a balance condition exists for our solution, namely, a restriction on the parameters that renders our solution conical-singularity-free. We have been careful to analyse regularity directly in $D=5$, since the analog $D=6$ analysis would not be conclusive: conically-singular $D$-dimensional metrics can uplift to completely regular $(D+1)$-dimensional ones via the KK formula (\ref{metric6d}).

First observe that the angular variables have periods $\Delta \phi = \Delta \psi = 2\pi$.  This simply follows from the required  asymptotically flat expression (\ref{AFmetric5}). The period of $\psi$ plays essentially no role in what follows, so we focus on $\phi$. While asymptotic flatness automatically ensures $\Delta\phi=2\pi$ along the rod $z\in(-\infty,a_1]$, the same is not guaranteed along the rod $z\in[a_2,a_3]$. The periodicity along this rod can be computed to be 
\be \label{PeriodCond}
\Delta\phi=2\pi\lim_{\rho\rightarrow0}\sqrt{\frac{\rho^2\, \tilde{G}_{ww}^{1/3} \,e^{2\nu}}{G^{(5)}_{ij}v^iv^j}} ,
\ee
where $v^i= (0,1,0)$ is the $D=5$ BZ vector for $z\in[a_2,a_3]$. Requiring (\ref{PeriodCond}) to match $\Delta\phi=2\pi$ yields the following constraint
\be\label{balance}
\frac{(a_3-a_4) \beta }{(a_3-a_2)   (\alpha +\beta-1 )\alpha}=1\,.
\ee
This balancing condition removes a disc of conical singularities inside  the black ring.
Just like the constraint~\eqref{constraint:a0}, this condition is valid whenever $\alpha\neq 1$ and $\beta\neq 0$. In the case $b_1=b_2=0$ the balance condition~\eqref{balance} should be replaced by
\be\label{balanceALT}
(a_3-a_0)(a_3-a_4)=(a_3-a_1)^2\,.
\ee

\subsubsection*{CTCs and Dirac-Misner strings}

Closed time-like curves (CTCs) tend to appear whenever solutions are over-spinning, as the supersymmetric case suggests (see e.g. \cite{Elvang:2005sa}). On practical grounds, CTCs will be absent if the angular submatrix $G^{(5)}_{ij}$ with $i,j=\phi,\psi$ is always positive semi-definite. Due to the complexity of the new solution, we have not analytically proved positiveness of that submatrix. However, we have performed extensive numerical tests confirming positiveness for wide coordinate ranges. As usually done in these cases, we take this as strong evidence that our solution is CTC-free. 

As for Dirac-Misner strings, these can be ruled out if the mixed metric component $G^{(5)}_{t\phi}$ vanishes at $\rho=0$ and $z=a_2,a_3$. Equivalently, as observed in \cite{Chen:2010ih}, these singularities can be ruled out in ISM constructions if  the rods $(-\infty,a_1]$ and $[a_2,a_3]$ do not have components along the timelike direction. From (\ref{LambdaConds}) we see that imposing $s_0=0$ we do obtain
\be\label{DMcons}
\Lambda_t\rightarrow 0\,.
\ee
Note that this condition aligns the rods $(-\infty,a_1]$ and $[a_2,a_3]$ of the uplifted 6D solution.

\subsubsection*{Parameter counting}

At this point we have concluded the regularity analysis of our solution. We will now track down the number of free parameters left by the imposition of all of the above constraints. Recall from section \ref{sec:DSCBR} that the ISM and the correction of the rod alignment produced a metric with 24 parameters. These we grouped into $(1)$ five rod endpoints ($a_0,a_1,a_2,a_3,a_4$), $(2)$ four BZ-parameters ($c_1, c_2, b_1, b_2$), and $(3)$ fifteen parameters, including $k$ and ($l_0, l_3$, $q_i,p_i,s_i)$, $i=0, \ldots ,3$.

First note that the ISM is insensitive to the absolute position of the rod configuration along the $z$-axis: it can be shifted along that axis without changing the solution. Accordingly, only the rod lengths $d_{ij}\equiv a_i-a_j$ are relevant. Only four of these are independent. The balance condition (\ref{balance}) gives one further constraint, thus leaving three free parameters within group (1). Without loss of generality, these can be taken to be $d_{31}, d_{32}, d_{34}$. Next we turn to group (2). The constraints (\ref{constraint:c1}), (\ref{constraint:c2}) and (\ref{constraint_b1b2}) leave only one of the BZ-parameters free, say $b_1$. Alternatively, the free parameter in this group can be taken to be $\beta$ defined in (\ref{def:alpha}). Finally, all parameters in group (3) are fixed: the asymptotically flat boundary condition imposed in Section \ref{section:sixfived} restricts 14 out of these 15 parameters. The remaining parameter in this group is fixed by the condition (\ref{DMcons}) for the absence of Dirac-Misner strings. Note that there are no further restrictions coming from the requirement of absence of CTCs.

As announced in section \ref{sec:DSCBR}, the regularity and asymptotic analyses leave only four parameters unfixed. Without loss of generality, these can be taken to be $(d_{31}, d_{32}, d_{34},\beta)$. Alternatively, as we choose in order to compute the extremal limit in section \ref{sec:PhysParam}, one can fix $a_3=1$ without loss of generality and work with independent parameters $(a_1, a_2, a_4, \beta)$. From the condition (\ref{balance}) we can fix
\bea \label{alphaCond}
\alpha=\frac{1}{2} \left(1-\beta +\sqrt{\frac{d_{32}(1-\beta)^2+4 d_{34} \beta}{d_{32}}}\right)\,.
\eea
Observe that the condition~\eqref{constraint:a0} and the ordering $a_0<a_1$ impose that the parameter $\beta$ in our solution ranges between certain intervals, depending on the relative locations of the soliton positions $a_i\,$: 
\bea
\left\{ 
\begin{array}{l}
0\leq\beta\le \frac{d_{32} \left(d_{31} d_{42} + d_{21} \sqrt{d_{31} d_{34}}\right)}{d_{31} \left(d_{32}^2-d_{31} d_{34}\right)} \qquad {\rm if} \quad  d_{32}^2-d_{31} d_{34} > 0\,, \\ 
\beta \geq 0 \qquad\qquad\qquad\qquad\qquad\qquad\; {\rm if} \quad  d_{32}^2-d_{31} d_{34} \leq 0 \,.
\end{array}
\right.
\eea
We have numerically checked that all physical parameters, that we present in the next section, are well defined for both ranges of $\beta$.

\section{The physical charges} \label{sec:PhysParam}

In this section we give the explicit expressions for the mass, angular momenta and electromagnetic charges supported by our solution.

\subsubsection*{Mass and angular momenta}

Using the coordinates $(r,\theta)$ introduced in (\ref{coordchange}), the mass $M$ and angular momenta $J_{\psi}$, $J_{\phi}$ can be read off from the next-to-leading order in the $r\rightarrow \infty$ asymptotic expansion of the full $D=5$ metric:
\be
ds_5^2\rightarrow \left(-1+\frac{8 M}{3\pi r^2}\right) dt^2-\frac{8\,J_{\psi} \sin^2\theta}{\pi r^2} dt\,d\psi-\frac{8\,J_{\phi} \cos^2\theta}{\pi r^2} dt\,d\phi+dr^2+r^2(d\theta^2+\sin\theta^2 d\psi^2+\cos\theta^2 d\phi^2)
\ee
Here, as well as throughout the remainder of the paper, we adopt natural units by setting the five-dimensional Newton constant and the speed of light to one, $G_5=c=1$.

Making use of the relations~\eqref{constraint:c1},~\eqref{constraint:c2} and~\eqref{constraint:b1b2}, but not imposing the constraint~\eqref{constraint:a0} nor the balance condition~\eqref{balance}, a calculation reveals the following expressions for the mass and the angular momenta:
\bea
M&=&\frac{\pi\,m}{4 d_{41} \left(d_{30}d_{32}(1-\beta )^2 - d_{10}d_{21}\alpha\beta \right)}\,,\\
J_{\psi}&=&\sqrt{\frac{d_{10} d_{20} d_{32}}{2}} \frac{\pi\,j_\psi}{d_{41}\left(d_{30} d_{32} (1-\beta)^2 - d_{10} d_{21} \alpha\beta \right)^{5/2}}\,,\\
J_{\phi}&=& \sqrt{\frac{d_{20} d_{21} d_{30} \beta}{2\alpha}} \frac{\pi\,j_\phi}{\left(d_{30} d_{32} (1-\beta)^2 - d_{10} d_{21} \alpha\beta \right)^{5/2}}\,,
\eea
where
\bea
m&=& - d_{10}d_{21} d_{31} d_{42} \alpha\beta - d_{30} d_{32} d_{41} \left[d_{42} (1+\beta)^2 - 2 d_{40} (1+\beta^2) + 3 d_{20} \beta^2 \right]\nonumber\\
&& + d_{20} d_{41} \left[d_{30} d_{32} + 3 d_{21} d_{31} \alpha\beta + d_{32}(d_{10}-2 d_{31}) \beta^2\right]\,,
\eea
\bea
j_{\psi}&=&d_{10}^2 d_{21} \alpha\beta^3 \left[d_{21} (d_{31} d_{41} - d_{40} d_{41} - d_{34} d_{42}) \alpha + d_{20} d_{32} d_{41} \beta\right] - d_{30}^3 d_{32}^2 d_{41} (1-\beta) \left(1-2\beta-\beta^3\right) \nonumber\\
&& -d_{30}^2 d_{32} \beta  \left\{d_{21} \left[ 2 d_{31} d_{41} \alpha (1-\beta)^3 - d_{40} d_{41} \alpha (1-\beta)^3 - d_{34} d_{42} \alpha (1-\beta)^3 + d_{32} d_{41} \beta \right]\right. \nonumber\\
&& \qquad \left.+ d_{10} d_{32} d_{41} \left(1-\beta +\beta^3\right) + d_{32} d_{41} \beta^2 (3 d_{32} (1-\beta) - \beta (d_{31} + (d_{20} - d_{31}) \beta))\right\} \nonumber\\
&& +d_{10} d_{30} \beta^2 \left\{2 d_{10} d_{21} d_{32} d_{41} \alpha (1-\beta) + d_{21} d_{32} \alpha (1-\beta) \left[d_{40} d_{41} + d_{34} d_{42} (1-\beta) - d_{41} d_{42} \beta\right] \right. \nonumber\\
&& \qquad \left.+ d_{21}^2 \alpha^2 \left[d_{34} (d_{21} + d_{42} \beta) + d_{40} d_{41} \beta + d_{31} (d_{41} - 2 d_{41} \beta)\right] - d_{20} d_{32}^2 d_{41} (1-\beta)^2 \beta \right\}\,,
\eea
\bea
j_{\phi}&=& -d_{30}^3 d_{32}^2 \alpha\beta^4 - d_{10}^2 d_{21} \alpha^2 \beta^2 \left[-d_{20} d_{21} \alpha - d_{31} (d_{32} + d_{21} \alpha ) + d_{32} (d_{42} + (2 d_{21}+d_{34}) \beta)\right]\nonumber\\
&&+d_{30}^2 d_{32} \left\{d_{32}^2 (1-\beta)^4 + d_{20} d_{21} \alpha^2 (1-\beta)^2 \beta\right.\nonumber\\
&& \qquad \left.+ d_{32} \alpha \left[d_{21} - d_{34} - (d_{21}-4d_{34}) \beta - 3(d_{21} + 2 d_{34}) \beta^2 + (5 d_{21} + 4 d_{34}) \beta^3 + (-2 d_{21} + d_{40}) \beta^4\right]\right\}\nonumber\\
&& -d_{10} d_{30} \alpha\beta \left\{d_{32}^2 d_{34} (1-\beta)^3 + d_{21}^2 \alpha (2 d_{32} + d_{20} \alpha\beta) \right.\nonumber\\
&& \qquad \left.+ d_{21} d_{32} \left[2 d_{32} (1-\beta)^3 + \alpha \left(d_{40} - (d_{10} + d_{20} + 2 d_{41}) \beta + (d_{10} + d_{42}) \beta^2\right)\right]\right\}\,.
\eea

We have also computed the angular velocities of the event horizon. Using condition~\eqref{constraint:a0} in order to eliminate $a_0$ from the expressions we obtain:
\bea
\Omega_\psi &=& \frac{-1}{d_{31} \alpha - d_{32} (\alpha+\beta-1)} \times \nonumber\\
&&\sqrt{\frac{\alpha (\alpha+\beta-1) (d_{31} \alpha \beta - d_{32} (\alpha+\beta-1)) \left[ d_{31} d_{32}(1-\beta)^2 - d_{21}(d_{31} \alpha \beta - d_{32} (\alpha+\beta-1)) \right]}{2\beta  \left({d_{31}}^2 \alpha \beta - {d_{32}}^2 (\alpha+\beta-1)\right)}} \,, \\
   \nonumber \\
\Omega_\phi &=& \sqrt{\frac{ \alpha(\alpha+\beta-1) \left[ d_{31}d_{32}(1-\beta)^2 - d_{21}(d_{31} \alpha \beta - d_{32} (\alpha+\beta-1)) \right]}{2 d_{21}d_{31}d_{32} \beta \left({d_{31}}^2 \alpha \beta -{d_{32}}^2 (\alpha+\beta-1)\right)}} \times \nonumber\\
   &&   \frac{ d_{31} d_{32} d_{41} (1-\beta) \beta - d_{21} \left[d_{32} d_{41} (\alpha+\beta-1) - d_{31} (d_{41} \alpha \beta + d_{32} (1-\beta) (\alpha+\beta-1))\right]}{ (d_{41}+d_{21} (\alpha+\beta-1)) (d_{31} \alpha - d_{32} (\alpha+\beta-1)) } \,.
\eea

\subsubsection*{Magnetic and electric charges} 

We now turn to the calculation of the electromagnetic charges carried by our solution. The (local) dipole charge is given by
\be
q\equiv\frac{1}{2\pi}\int_{S^2} F\, .
\ee
Recall that $F=dA$ is the abelian two-form field strength of the  Maxwell field, and the $S^2$-sphere supporting the integration is parametrized by $(z,\phi)$ at constant $t,\rho,\psi$. For our solution, displaying rotation in both directions and non-trivial electric charge, there is an electric moment induced by the rotation of the electric charge, which makes the calculation quite subtle. The final expression for the magnetic dipole charge, employing condition~\eqref{constraint:a0} to eliminate $a_0$, is
\be
q\equiv \frac{\Delta\phi}{2\pi}\lim_{\rho\rightarrow0}\left[A_{\phi}|_{z\in(-\infty,a_1]}-A_{\phi}|_{z\in[a_2,a_3]}\right]=\sqrt{\frac{2d_{31}d_{32}d_{42}}{d_{34} d_{41} }}\, \frac{d_{21} \alpha +d_{42}(1- \beta)}{\left(d_{31}d_{32} (1-\beta )^2 - d_{21} \left[d_{31}\alpha\beta - d_{32} (\alpha +\beta -1)\right]\right)^{1/2}}\,.
\ee
In the Pomeransky-Sen'kov limit, $a_4\to a_2$, the dipole charge does vanish, as expected. Taking instead the single rotation limit, $\alpha\to1$ and $\beta\to0$, we precisely reproduce the dipole charge of the singly spinning dipole black ring~\cite{Emparan:2004wy}.

The electric charge can be read off from the asymptotic behaviour of the $A_t$ gauge field component,
\be
A_t=-\frac{Q}{2 r^2}+O[1/r^3]\, .
\ee
After some calculation, we find
\be
Q\equiv\frac{1}{2\pi}\int_{S^3} \star F= 4\sqrt{\frac{d_{20}d_{21}d_{32}d_{34}d_{42}\,\alpha\beta}{d_{41}}}\, \frac{d_{31}\beta - d_{30}}{d_{30}d_{32}(1-\beta )^2 - d_{10}d_{21}\alpha\beta}\,,
\label{electric_charge}
\ee
where the integral is taken over a spatial 3-sphere of infinite radius.
As a consistency check, we have also verified that this expression vanishes, as expected, both when taking the neutral, Pomeransky-Sen'kov \cite{Pomeransky:2006bd} limit, $a_4\to a_2$,  and also when turning off the rotation on the $S^2$, as in \cite{Emparan:2004wy}.

\subsubsection*{Horizon area, temperature and extremal limit}

The event horizon area can be computed directly in Weyl coordinates as the 3-volume determined from the metric induced on the timelike rod, at constant $t$. The temperature, on the other hand, may be obtained by requiring the absence of a conical singularity at the horizon in the Wick-rotated metric~\cite{Harmark:2004rm}.
Performing these calculations we arrive at
\bea
A_{H} &=& 4 \pi ^2 d_{21} \sqrt{\frac{2\, d_{20}\, d_{32}  \left(\beta\, d_{10}+(1-\beta )\, d_{30}\right)^2 \left(\alpha \beta \, d_{21} d_{31}{}^2+d_{30} \, d_{32} \, d_{41}\right)^2}{d_{31}{}^2\, d_{41} \left((1-\beta )^2 d_{30} d_{32}-\alpha  \beta \, d_{10}\, d_{21}\right){}^3 }}\,, \label{area}\\
 T_H &=& 2\pi\,d_{21}/A_H\,. \label{temperature}
\eea
We have further verified that the neutral limit, $a_4\to a_2$, of these expressions correctly reproduces the physical quantities of the unbalanced Pomeransky-Sen'kov black ring as given in~\cite{Chen:2011jb}. Similarly, in the single rotation limit, $\alpha\to1$ and $\beta\to0$, we recover the thermodynamic quantities of Emparan's dipole black ring~\cite{Emparan:2004wy}.\footnote{This comparison requires a careful rescaling because, unlike us, Ref.~\cite{Emparan:2004wy} adopts a normalization for the angular coordinates such that their periodicities are non-standard, $\Delta\phi,\Delta\psi\neq2\pi$.}

A careful analysis reveals the existence of a well-behaved zero temperature limit for our solution, where all physical quantities remain finite.  As first noticed in \cite{Elvang:2007hs}, the extremal, zero-temperature limit corresponds to a collapse of the rods. In order to take this limit, we can choose $a_3=1$ without loss of generality, and rescale the remaining five parameters (note that one of them, say $\alpha$, depends on the other four, see above equation (\ref{alphaCond})) as
\be
a_1= w_1 \,\epsilon\,,\qquad a_2= w_2 \, \epsilon\,,\qquad a_4= w_4\, \epsilon\,,\qquad \alpha=1+w_{\alpha}\, \epsilon\,,\qquad \beta=1+w_{\beta}\, \epsilon\,.
\label{extremal_limit}
\ee
where $0<w_1<w_2<w_4<1$. In the limit $\epsilon \rightarrow 0$, the temperature (\ref{temperature}) goes to zero linearly and the area (\ref{area}) reduces to the finite expression
\be
A_H=8 \pi ^2 \sqrt{\frac{(w_2-w_1)^3(w_4+w_2-2 w_1) ^2}{(w_1-w_4) (w_1-w_2-w_{\beta})^3 (w_1-w_2+w_{\beta} )}}\,.
\ee
We have checked that the angular momenta $J_{\phi},J_{\psi}$ and charges $q, Q$ remain finite in this limit. It would be interesting to further investigate BPS limits of our solution.

\subsubsection*{Phase diagrams}

\begin{figure}[t!]
\centering{
\begin{tabular}{lr}
\includegraphics[width=8.5cm]{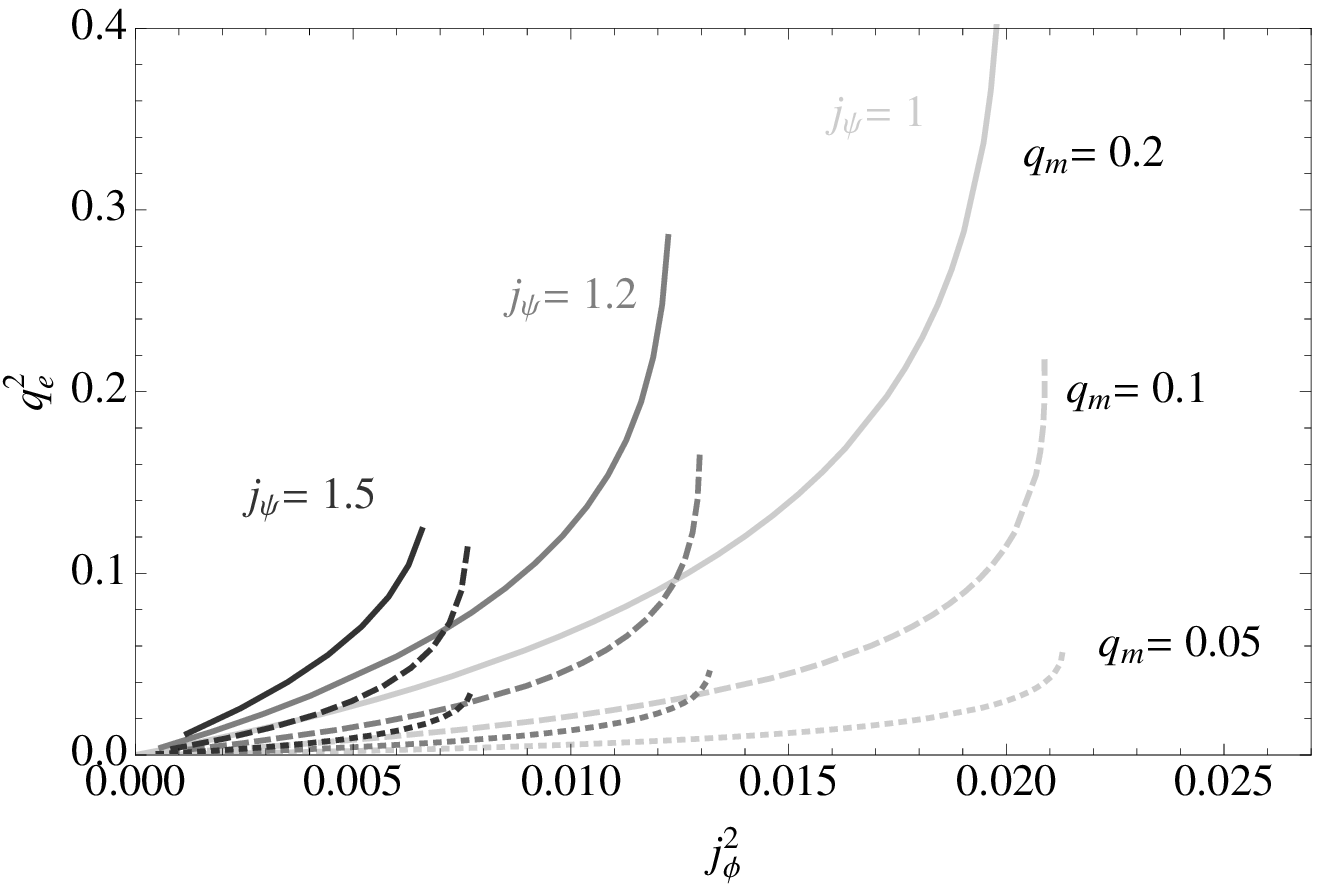}
&
\includegraphics[width=8.5cm]{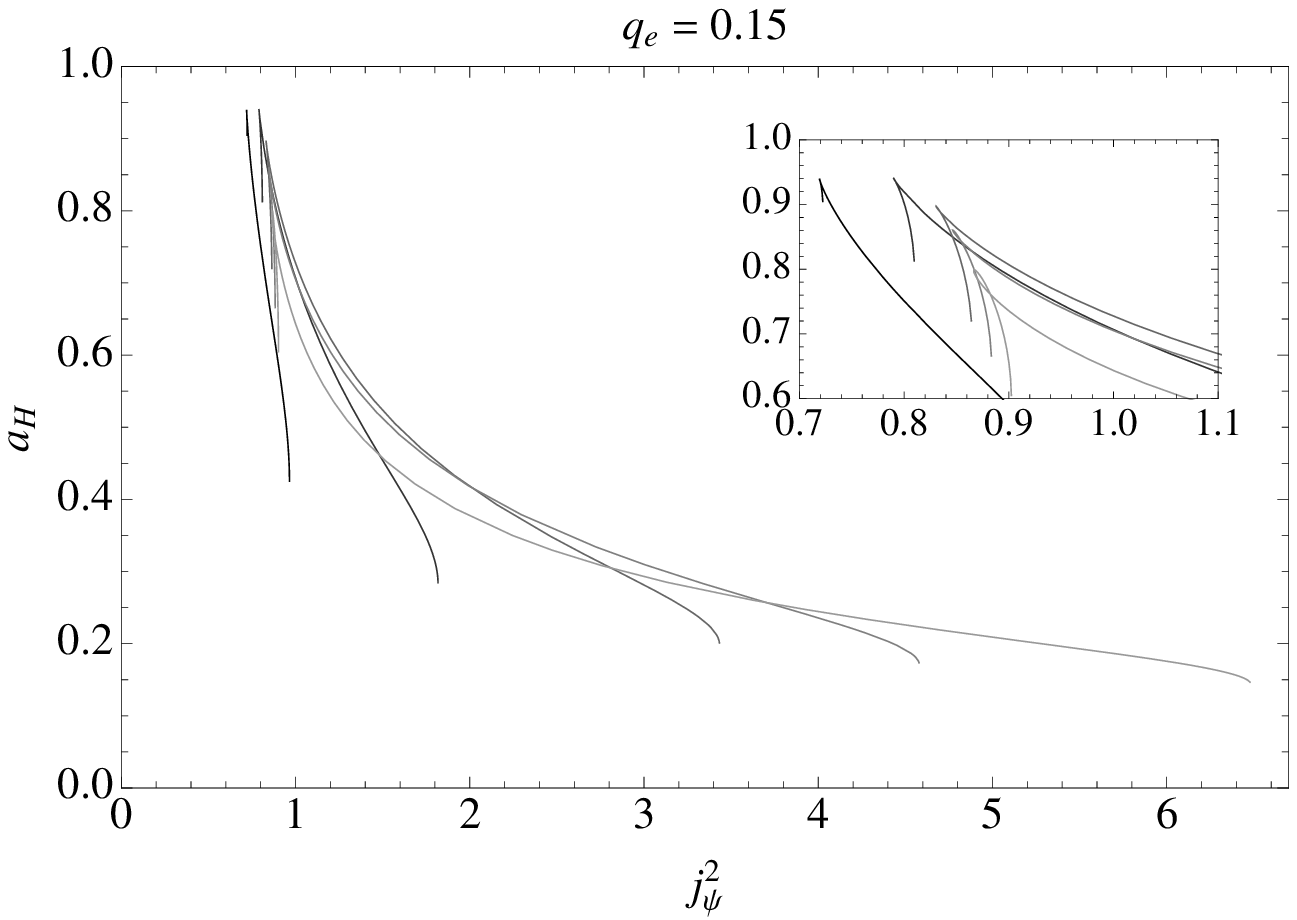}
\end{tabular}}
\caption{(a) The phase diagram on the left panel represents $(j_\phi^2,q_e^2)$ for fixed values of $j_\psi$ and $q_m$. Light-grey curves correspond to $j_\psi=1$, medium-grey to $j_\psi=1.2$ and dark-grey to $j_\psi=1.5$. Solid line curves correspond to $q_m=0.2$, dashed lines to $q_m=0.1$ and dotted lines to $q_m=0.05$.
(b) The phase diagram on the right panel represents $(j_\psi^2,a_H)$ for a fixed value of $q_e=0.15$ and five different fixed values for $j_\phi$. The different shades of grey correspond (from darkest to lightest) to $j_\phi=0.15, 0.10, 0.07, 0.06, 0.05$. The inset  zooms in the region where the curves attain maximum horizon area. Note that $q_m$ varies along each of these curves.}
\label{phasediags1}
\end{figure}

\begin{figure}[t]
\centering{
\begin{tabular}{lr}
\includegraphics[width=8.5cm]{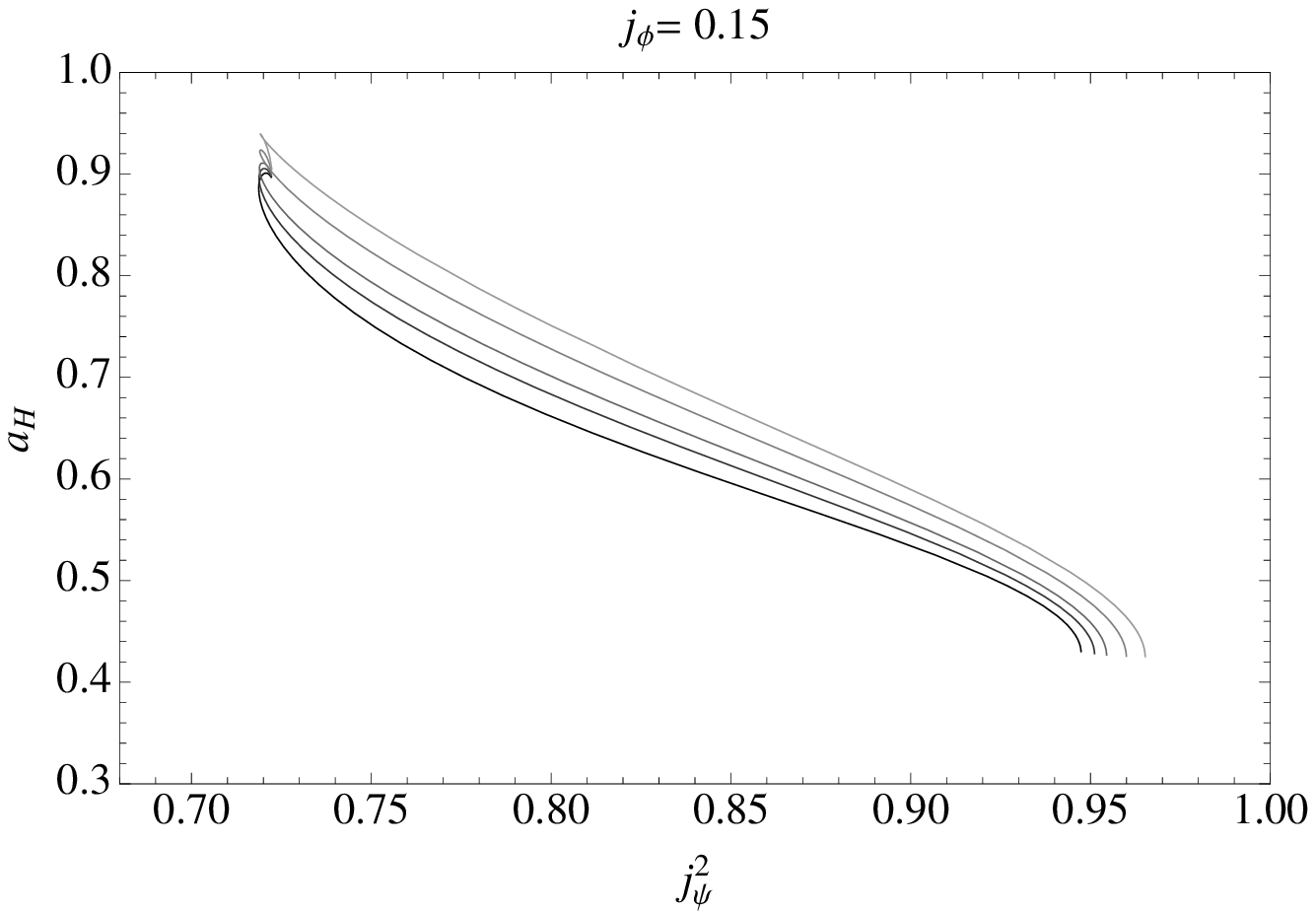}
&
\includegraphics[width=8.5cm]{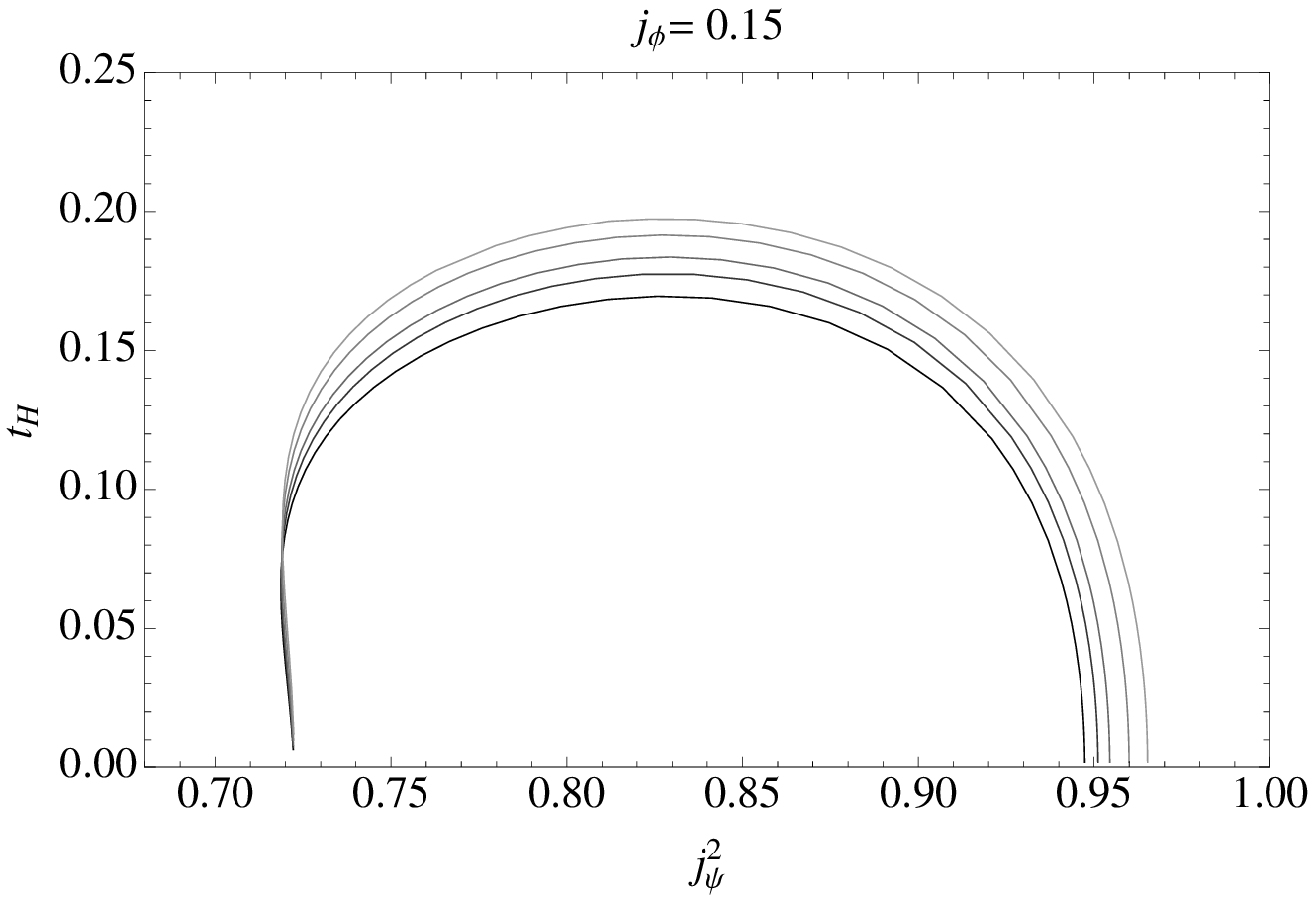}
\end{tabular}}
\caption{Phase diagrams for a fixed value of $j_\phi=0.15$ and five different fixed values for $q_e$. The different shades of grey correspond (from darkest to lightest) to $q_e=0.50, 0.45, 0.40, 0.30, 0.15$. (a) The left panel represents $(j_\psi^2,a_H)$. (b) The right panel shows $(j_\psi^2,t_H)$. Note that $q_m$ varies along each of these curves.}
\label{phasediags2}
\end{figure}

Having obtained expressions for the relevant physical quantities describing our family of black rings, we can now study the parameter space they cover in terms of phase diagrams. Given the complexity of the solution and the large number of charges it carries, there are various possible phase diagrams one can consider. The theory under consideration -- being obtained from six dimensional vacuum gravity by KK reduction -- is scale invariant. To remove this rescaling freedom it is common to define dimensionless combinations for the charges,
\be
j_\psi^2 = \frac{27\pi}{32} \frac{J_\psi^2}{M^3}\,,\qquad\qquad
j_\phi^2 = \frac{27\pi}{32} \frac{J_\phi^2}{M^3}\,,\qquad\qquad
q_e^2 = \frac{Q^2}{M}\,,\qquad\qquad
q_m^2 = \frac{q^2}{M}\,,
\label{dimless_charges}
\ee
and also for the event horizon area and Hawking temperature,
\be
a_H = \frac{3}{16}\sqrt{\frac{3}{\pi}} \frac{A_H}{M^{3/2}}\,, \qquad\qquad t_H = \sqrt{\pi}M^{1/2}T_H\,.
\label{dimless_area_temp}
\ee
The phase space may then be characterized by five parameters, $(j_\psi, j_\phi, q_e, q_m, a_H)$.  In Figures~\ref{phasediags1} and \ref{phasediags2} we display the phase diagrams for a few representative choices of fixed dimensionless charges. We now comment on some relevant features.

Recall that our solution carries five charges (mass, two angular momenta, electric charge and magnetic dipole charge) but only four are independent. Correspondingly, out of the four dimensionless combinations~\eqref{dimless_charges} only three are independent. Hence, a phase diagram plotting $a_H$ (or $t_H$) vs $j_\psi$, with other two dimensionless combinations fixed (say, $j_\phi$ and $q_e$), implies that the remaining combination (in this case it would be $q_m$) will also vary along the curve. This constraint is clearly evident in the left panel of Fig.~\ref{phasediags1}, where we fix $j_\psi$ and $q_m$ and plot how $q_e^2$ varies with $j_\phi^2$. It is also apparent that when rotation on the $S^2$ is turned off the electric charge vanishes, in agreement with the discussion below Eq.~\eqref{electric_charge}.
The right panel of Fig.~\ref{phasediags1} displays the phase diagram $(j_\psi^2,a_H)$ for a fixed finite value of the electric charge $q_e=0.15$ and several fixed values for the reduced $S^2$ angular momentum $j_\phi$.

Fig.~\ref{phasediags2} exhibits the impact of the electric charge on the phase diagrams for our family of solutions. For fixed $j_\phi$ and $j_\psi$ an increase in $q_e$ typically leads to a decrease both in the reduced horizon area $a_H$ and in the reduced temperature $t_H$. One final comment concerns the extremal limit. The regular behaviour of the $\epsilon\to0$ limit, in the parametrization of~\eqref{extremal_limit}, is apparent in Fig.~\ref{phasediags2}: the right panel of that figure shows that the temperature goes to zero in this limit, while the left panel confirms that the horizon area remains finite. The extremal limit we discussed in the previous subsection actually corresponds to the right-end points of the curves in Fig.~\ref{phasediags2}. Based on our experience with doubly-spinning (but neutral) black rings~\cite{Elvang:2007hs}, the left-end of these curves should correspond to the collapse of the black rings to black holes. This is confirmed by the fact that the length of the rod $z\in[a_2,a_3]$ shrinks to zero in such limit.

\section{Conclusions} \label{sec:conclusions}

In this paper we have presented the details of the construction of a new black ring in $D=5$ Einstein-Maxwell-dilaton theory with Kaluza-Klein dilaton coupling. Guided by the insight gained in \cite{Rocha:2011vv}, we have relied on heavy use of inverse-scattering techniques in six dimensions in order to construct a new $D=6$ Ricci-flat solution that KK reduces to a $D=5$ doubly spinning, electrically charged version of the ring of \cite{Emparan:2004wy}. Convenient parameter tuning determined by regularity and asymptotic analyses renders a solution free of conical singularities, CTCs and Dirac-Misner strings, depending on four independent parameters.
These are related to the five physical charges (subject to one constraint) that our black ring carries: mass $M$, angular momenta $J_\psi$ and $J_\phi$ along the $S^1$ and $S^2$ of the near horizon topology, and electric charge $Q$ and magnetic dipole charge $q$ with respect to the abelian vector field $A$ present in the theory.

The (unbalanced version of our) novel, exact black ring solution is the most general black ring constructed to date in a minimal extension of $D=5$ GR. Based on arguments similar to those of \cite{Elvang:2004xi}, the most general non-supersymmetric black ring solution to $D=5$ Einstein-Maxwell-dilaton theory should be expected to depend on five independent parameters, related to completely unconstrained physical charges $(M,J_\psi, J_\phi, Q,q)$. The solution in appendix \ref{app}, containing an additional free parameter, should correspond to this five-parameter generalization, as our preliminary studies suggest.

It would be interesting to extend the construction of charged black rings to more general theories than just Einstein-Maxwell-dilaton. Further studying the interplay of the ISM method with other integrability techniques, like the solution generating method based on U-dualities, along the lines of  \cite{Figueras:2009mc}, might be very useful in this direction. An especially interesting theory to consider would be $U(1)^3$ supergravity, where the most general black ring should contain nine parameters \cite{Elvang:2004xi}. String \cite{Elvang:2004xi,Hoskisson:2008qq} and supergravity \cite{Gal'tsov:2009da} dualities have been employed to obtain charged rings in that theory. Setting several charges equal, charged solutions in Einstein-Maxwell-dilaton can be obtained \cite{Gal'tsov:2009da}, making contact with the solutions of this paper.

Finally, it has been recently pointed out that the inner horizon of non-extremal black holes might play a relevant role in the determination of their still elusive microscopics. Building on some earlier results, it has been recently shown that the products of the areas of the inner and outter horizons of spherical non-extremal black holes is a polynomial on the electromagnetic charges and angular momenta, and is independent of the mass or possible moduli \cite{Cvetic:2010mn}. This result has been extended for black rings of several theories in \cite{Castro:2012av}. It would be interesting to establish the universality of the product area law  for our solutions.

\subsection*{Acknowledgements}

 We are grateful to Roberto Emparan for useful discussions. We also thank Amitabh Virmani for collaboration during the early stages of this work. J.~V.~R. is supported by {\it Funda\c{c}\~ao para a Ci\^encia e Tecnologia} (FCT)-Portugal through contract no. SFRH/BPD/47332/2008. M.J.R. is supported by the European Commission - Marie Curie grant PIOF- GA 2010-275082. O.V. is supported in part by the Netherlands Organisation for Scientific Research (NWO) under the VICI grant 680-47-603, and by the Spanish Government research grant FIS2008-01980.

\appendix
\section{ Full five-parameter solution} \label{app}

In section \ref{sec:IMS} we described in detail the inverse scattering transformations that generate (the six-dimensional lift of) a  general $D=5$ Einstein-Maxwell-dilaton black ring. For simplicity, in the remainder of the paper we set the BZ parameter $b_3$ to zero. We wrote the corresponding solution and analysed in detail the regularity and physical charges of the resulting four-parameter black ring. In this appendix, we recover a non-vanishing $b_3$ parameter to write an even more general solution, which thus generalises that of section \ref{sec:DSCBR}. We have performed a full regularity analysis of this more general solution, which it successfully passes. We will however omit the details, as the analysis of section \ref{subsec:RodStr} remains mostly unchanged: only eqs.~\eqref{p3}, \eqref{LambdaConds} and~(\ref{Omega2})--(\ref{OmegaD}) are affected by the inclusion of additional terms that are present only for $b_3\neq0$. In particular, the constraints~\eqref{constraint:c1}, \eqref{constraint:c2} and~\eqref{constraint_b1b2} still apply. We can conjecture that the solution that we now present, having one more parameter than the solution discussed in detail in the bulk of this paper, in fact is the most general black ring in Einstein-Maxwell-dilaton with KK coupling. We only give here the solution. The calculation of the physical charges will be presented elsewhere.\footnote{We have explicitly checked that the five parameter solution of this appendix is curvature-singularity free: the Kretschmann invariant (the square of the Riemann tensor) is everywhere smooth. This solution thus cures a possible Kretschmann divergence at $(\rho,z)=(0,a_2)$ of our four-parameter solution, corresponding to the inner pole of the $S^2$ component of the ring-like event horizon. We thank the authors of~\cite{Chen:2012kd} for pointing out this possible singularity.}

The metric containing the additional parameter $b_3$ is
\be
ds^2_6 = (G)_{ab}\; dx^a dx^b+ e^{2\nu}(d\rho^2+dz^2),\qquad x^a=(t, \phi, \psi, w)\,,
\label{newmetric}
\ee
where
\be
G = (G'_0-N_v^T\,\Gamma^{-1}\,N_v)\,\zeta^{-1}, \qquad e^{2\nu}=e^{2\nu_0} \frac{\det\Gamma}{\det\Gamma_0}\,.
\label{newGconf}
\ee
Here,  $(G'_0,e^{2\nu_0})$ are again the seed metric expressions given in \eqref{seed1} and \eqref{e2nu}, and $\zeta^{-1}=\mu_1 \mu_4/\rho^2$ was defined in section \ref{sec:IMS}. The determinant of $\Gamma_0$ is
\be
\det\Gamma_0=-\frac{ \mu_1 \mu_2 \mu_3^5 \mu_4 (\mu_1-\mu_2)^2 (\mu_3-\mu_4)^2 Z_{00} Z_{12}^2 Z_{22}^3 Z_{34}^2}{\rho^4 \mu_0  (\mu_0-\mu_3)^2 (\mu_1-\mu_3)^2 (\mu_2-\mu_3)^4 (\mu_2-\mu_4)^2 Z_{04}^2 Z_{11} Z_{13}^4 Z_{14}^2 Z_{23}^2 Z_{44} }\,,
\ee
where we have introduced the function $Z_{ij}=\rho^2+\mu_i\mu_j$. The remaining quantities that appear in (\ref{newGconf}) are the matrices $N_v$ and $\Gamma$. Their explicit $(\rho,z)$-dependent expressions are 
\be 
N_v=
 \begin{pmatrix}
   \frac{c_1 \rho^2 \left(\rho^2+\mu_0^2\right) (\mu_2-\mu_0)}{\mu_0 \left(\rho^2+\mu_0 \mu_1\right) \mu_2 \left(\rho^2+\mu_0 \mu_4\right)} & 0& \frac{\left(\rho^2+\mu_0^2\right) \mu_3}{\mu_0 (\mu_3-\mu_0) \left(\rho^2+\mu_0 \mu_4\right)} & 0\\
   0 & \frac{\rho^4 (\mu_1-\mu_2)^2 \mu_3}{\mu_1 \mu_2^2 (\mu_3-\mu_1) \left(\rho^2+\mu_1 \mu_3\right)^2} & \frac{b_1 \left(\rho^2+\mu_0 \mu_1\right) \mu_3}{\mu_1 (\mu_3-\mu_1) \left(\rho^2+\mu_1 \mu_4\right)} &0\\
    \frac{b_2 (\mu_0-\mu_2) \left(\rho^2+\mu_2^2\right)}{\rho^2 (\mu_2-\mu_1) (\mu_2-\mu_4)} & -\frac{\left(\rho^2+\mu_2^2\right)^2 \mu_3}{\rho^2 (\mu_2-\mu_3)^2 \left(\rho^2+\mu_2 \mu_3\right)} & 0 & \frac{b_3 \mu_2^2(\mu_2-\mu_3)(\mu_2\mu_3+\rho^2)}{\rho^2\mu_3(\mu_1-\mu_2)(\mu_2^2+\rho^2)}\\
   0 & \frac{c_2 \rho^4 \mu_3 (\mu_2-\mu_4)^2}{\mu_2^2 (\mu_3-\mu_4) \mu_4 \left(\rho^2+\mu_3 \mu_4\right)^2} & 0 &\frac{\mu_2 (\mu_3-\mu_4) \left(\rho^2+\mu_3 \mu_4\right)}{\mu_3 (\mu_2-\mu_4) \mu_4 \left(\rho^2+\mu_1 \mu_4\right)}\\
 \end{pmatrix}
\ee
and
\be
\Gamma=
 \begin{pmatrix}
   \Gamma_{11} & \Gamma_{12}& \Gamma_{13} & 0\\
   \Gamma_{12} & \Gamma_{22} & \Gamma_{23} &\Gamma_{24}\\
   \Gamma_{13} & \Gamma_{23} & \Gamma_{33} &\Gamma_{34}\\ 
   0 & \Gamma_{24} & \Gamma_{34} &\Gamma_{44}\\
 \end{pmatrix}\,,
\ee
with components
\be \label{GammaComps}
\Gamma_{11} = \frac{-c_1^2 \rho^2 \mu_1 \mu_4 (\mu_0-\mu_2)^2 (\mu_0-\mu_3)^2 Z_{00} - \mu_2 \mu_3 \mu_4 Z_{00} Z_{01}^2}{\mu_0 \mu_2 (\mu_0-\mu_3)^2 Z_{01}^2 Z_{04}^2}\nonumber\,,\\
\ee
\be
\Gamma_{12} = \frac{b_1 \mu_3 \mu_4 Z_{00} }{\mu_0 (\mu_0-\mu_3) (\mu_3-\mu_1) Z_{04}Z_{14}}\nonumber\,,\qquad \Gamma_{13}=\frac{b_2 c_1 \mu_1 \mu_4 (\mu_0-\mu_2) Z_{00} Z_{22} }{\rho^2 \mu_0 (\mu_2-\mu_1) (\mu_2-\mu_4)Z_{01} Z_{04}}\nonumber\,,
\ee
\be
\Gamma_{22}= \frac{-b_1^2  \mu_2^2 \mu_3  \mu_4 Z_{01}^2 Z_{13}^4+\rho^4 \mu_0 \mu_3^3 (\mu_1-\mu_2)^4  Z_{14}^2}{\mu_0 \mu_2^2 (\mu_1-\mu_3)^2 Z_{11} Z_{13}^4 Z_{14}^2}\,, \nonumber
\ee
\be
\Gamma_{23}= \frac{ \mu_3^3 (\mu_1-\mu_2) Z_{22}^2}{\rho^2 (\mu_1-\mu_3) (\mu_2-\mu_3)^2 Z_{13}^2 Z_{23}}\,,\qquad \Gamma_{24}= -\frac{c_2 \rho^4 \mu_3^3 (\mu_1-\mu_2)^2(\mu_2-\mu_4)^2}{\mu_2^2 (\mu_1-\mu_3) (\mu_3-\mu_4) Z_{13}^2 Z_{14} Z_{34}^2}\,,
\ee
\be
\Gamma_{33}= \frac{\mu_2 \mu_3^2 Z_{22}^4 \left(\mu_0 \mu_2 \mu_3^3 (\mu_1-\mu_2)^2 (\mu_2-\mu_4)^2 Z_{22}^2 - b_2^2 \rho^2 \mu_1 \mu_4 (\mu_0-\mu_2)^2 (\mu_2-\mu_3)^4 Z_{23}^2 \right)-Z}{\rho^6 \mu_0 \mu_3^2 (\mu_1-\mu_2)^2 (\mu_2-\mu_3)^4 (\mu_2-\mu_4)^2 Z_{22}^3 Z_{23}^2}\,, \nonumber
\ee
\be
\Gamma_{34}= \frac{c_2 \mu_3^5(\mu_1-\mu_2) (\mu_2-\mu_4)^3 Z_{22}^3 Z_{14}+b_3\mu_1\mu_2^2(\mu_2-\mu_3)^3(\mu_3-\mu_4)^2 Z_{23}^2 Z_{34}^3 }{\rho^2\mu_3^3(\mu_1-\mu_2) (\mu_2-\mu_3)^2(\mu_2-\mu_4)^2(\mu_3-\mu_4) Z_{22} Z_{23} Z_{14} Z_{34}^2}\,,\nonumber
\ee
\be
 \Gamma_{44}= \frac{c_2^2 \rho^4 \mu_3^5 (\mu_2-\mu_4)^6 Z_{14}^2-\mu_1 \mu_2^3 (\mu_3-\mu_4)^4 Z_{34}^6}{\mu_2^2 \mu_3^2 (\mu_2-\mu_4)^2 (\mu_3-\mu_4)^2  Z_{14}^2 Z_{34}^4 Z_{44}}\,.\nonumber
\ee
where $Z=b_3^2 \rho^4\mu_0\mu_1\mu_2^3(\mu_2-\mu_3)^6(\mu_2-\mu_4)^2 Z_{23}^4$.

The corresponding $D=5$ Einstein-Maxwell-dilaton solution, leading to a five-parameter black ring once the regularity and asymptotic conditions are imposed, is finally obtained upon substitution of the above expressions in the reduction furmulae (\ref{5dsol}), (\ref{5dsol2}).


\end{document}